**Neural Hardware for the Language of Thought: New Rules for an Old Game**[1]

Gualtiero Piccinini (University of Missouri)

Abstract: The Language of Thought (LOT) hypothesis posits that at least some important cognitive processes involve language-like representations. These representations must be processed by appropriate hardware. Since the organ of biological cognition is the nervous system, whether biological cognition relies on a LOT depends on how neural hardware works. I distinguish between different versions of LOT, articulate their hardware requirements, and consider which versions of LOT are supported by empirical evidence. I argue that the Classical LOT hypothesis (Fodor 1975) is ruled out; the version of LOT that is best supported by empirical evidence is the Nonclassical LOT thesis that some neural representations mirror some of the structure of natural language and represent in a language-like way, yet they encode information nondigitally and are processed by ordinary (nondigital, and hence Nonclassical) neural computations that rely not only on syntactic structure but many other features.

*This report is a revised version of an unpublished essay that has circulated since early 2025. Since it has been cited in print and others are engaging with it in unpublished manuscripts, I am making it public in this form while I work on a monograph that will draw from this material.*

---

[1] This work was partially done on the land of the Osage Nation, Otae-Missouri, Chikasaw, Illni, Ioway, Quapaw, Shawnee, Delaware, Kickapoo, Sac & Fox, Omaha, and Santee Sioux. This paper builds on 30+ years of research and too many conversations for me to thank everyone involved; among the most relevant, memorable, and helpful are exchanges with Hessameddin Akhlaghpour, Sonya Bahar, Kenneth Black, Trey Boone, Robert Brandom, Matt Brown, David Chalmers, David Colaço, Matteo Colombo, Guy Dove, Frankie Egan, the late Jerry Fodor, Randy Gallistel, David Glanzman, the late Gilbert Harman, John Krakauer, Corey Maley, Eric Margolis, Brian McLaughlin, Marcin Milkowski, Alex Morgan, Mirinda James, Michael Kremer, Joe Lau, John McDowell, Jonathan Najenson, Tomaso Poggio, Kyrill Potapov, Michael Rescorla, Brendan Ritchie, David Rosenthal, Carl Sachs, Richard Samuels, Steve Selesnik, Mark Sprevak, Wayne Wu, and two anonymous readers. Many thanks to Hanzhe Dong and Omar Ghaffar for invaluable research assistance and illuminating exchanges. Special thanks to David Barack for detailed and very helpful feedback on the earliest drafts of this essay. Thanks to the many audiences who heard some of this material for their attention and feedback.

## 1. New Rules for an Old Game

The Language of Thought (LOT) hypothesis holds that at least some important cognitive processes involve "language-like" (Fodor, Bever, and Garrett 1974) representations that constitute a "mentalese" (Sellars 1964) or "language of thought" (Harman 1968a, b, 1969, 1970a, b).[2] In LOT's most influential version (Fodor 1968, 1972, 1975; Fodor, Bever, and Garrett 1974), mentalese is a formal language consisting of language-like data or instructions processed by a computing system with an architecture similar to that of ordinary digital computers ("Turing/von Neumann architectures," writes Fodor 1987, 139). I will refer to this as *Classical LOT*.

Classical LOT has been called the "only" or "best" game in town. But the game's rules were always too fuzzy to determine a winner. In this paper, I propose rules that are clear and cogent enough to play fairly and take the game to the next level. Following these improved rules, I will argue that Classical LOT is not a viable endgame. If you are interested in propositional thought, the endgame involves a kind of *Nonclassical* LOT I will sketch. In brief, my argument is that any LOT hypothesis, Classical or Nonclassical, requires hardware with the ability to process language-like representations. Since biological cognition is carried out by nervous systems, any plausible LOT hypothesis must be consistent with how nervous systems work. And we know enough about neural hardware to perform an eliminative induction against Classical LOT in favor of Nonclassical LOT.

Here is how to play. By "cognitive processes" (or "thought"), I mean reasoning, planning, imagining, decision making, and so forth. By "hardware", I mean the (implemented) components that process representations, such as microchips within digital computers and biophysical neurons within brains.[3] By "representation", I mean states or state sequences—such as strings of digits within a digital computer or spike trains within a nervous system—that carry semantic content. Since the representations in question are realized in nervous systems, I will follow the mainstream and refer to them as *neural representations*—representations of the sort that is observed in nervous systems (Thomson and Piccinini 2018). By the same token, I will refer to computations over neural representations as *neural computations*.

Crucially, neural representations and computations have *compositional* structure such that simpler, lower-level representations (e.g., spike trains from single neurons) compose more complex, higher-level representations (e.g., neural manifolds from neuronal assemblies or populations), computational operations on simpler representations compose computational operations on more complex representations, and the semantic content of composite representations is largely a function of the semantic content of their component representations plus the way they compose. For instance, what the cortical visual system represents and computes is largely a function of what cortical visual areas represent and compute, which is largely a function of what individual cortical columns within each area represent and compute, which in turn is largely a function of what neurons within each column represent and compute.[4]

---

[2] The term "language of thought" was used at least since the 19th century, often to mean something like Leibniz's characteristica universalis (e.g., Gadamer 1967). Harman appears to be the first who used it in the sense relevant to this essay.

[3] Sometimes, nervous systems are said to contain "wetware" rather than "hardware", to stress that neural tissues are alive, more plastic than typical computer hardware, and bathed in blood, interstitial and cerebrospinal fluids, and a soup of neurotransmitters, hormones, and other signaling biomolecules. For present purposes, neural wetware is a type of hardware.

Hardware constrains and is constrained by *computational architecture*, which is the system of organized computing components that process representations appropriate for that hardware in accordance with algorithms appropriate for that hardware. For example, a microchip within a digital computer may play the computational role of *central processing unit*, or a neural circuit might carry out the computational operation of *normalization* (Carandini and Heeger 2012). As I will discuss in more detail shortly, representations cannot be processed, and algorithms cannot be followed, unless an appropriate computational architecture is in place. Therefore, the relation between hardware, architecture, algorithms, and representations is central to a proper assessment of any theory of cognition such as LOT.

The importance of computational architecture has been underappreciated, perhaps in part because Marr (1982) omits it from his influential framework for analyzing computing systems. Marr articulates three levels of analysis: computational (e.g., multiplication), algorithmic (e.g., multiplying by computing partial products and then summing them), and implementation (e.g., a microchip). He skips computational architecture (e.g., the CPU and memory of a von Neumann architecture implemented by microchips), which lies between the algorithmic and implementation levels. Algorithms run on a computational architecture, which is realized by hardware and explains how the hardware can process relevant representations in accordance with relevant algorithms. In what follows, I will consider four levels of analysis: computational, algorithmic, architecture, and hardware. This four-level framework matches Marr and Poggio's original framework (1976), the difference being that Marr and Poggio use the term "mechanisms" where I use "architecture".[5]

---

[4] Semantic compositionality in neural representations, unlike semantic compositionality in formal logic, is not always transparently intelligible to external observers. Some neural manifolds may have semantic content that is difficult to decompose in easily interpretable ways (cf. Burnston 2021). More generally, one and the same multilevel phenomenon, such as neural computation, often depends on complex relationships between processes that occur at different scales (cf. Rice 2024). A detailed account goes beyond the scope of this article. For recent advances in understanding neural representations and their content, see Nestor 2024, Heemskerk 2025, and Martinez 2025. For a defense of the view that neural representation and computation are multilevel, see Counts 2025. For an argument that the LOT hypothesis needs to be integrated with cognitive neuroscience, see Schneider 2011. For a defense of the relevant integrationist framework, including critiques of assumptions sometimes fallaciously invoked in defense of Classical LOT, such as the alleged autonomy of psychology and the Church-Turing thesis, see Piccinini and Craver 2011; Morgan and Piccinini 2018; Piccinini 2020a, 2020b, 2022; Piccinini and Hetherington 2025; Piccinini and Fuentes 2025.

[5] The "Marrian" three-level framework was already outlined by Reichardt and Poggio (1976). According to Poggio (pers. comm.), Marr (1982) omitted the architectural level to keep the framework simple given that in the nervous system you must study the one and only implementation to understand its architecture; another consideration was that the three-level framework matched the three levels described by Reichardt and Poggio (1976). For our purposes, it is critical to understand the relations between *all* four levels of analysis. Computational architecture, as

I will advance the debate on LOT by taking three important steps. In Section 2, I distinguish different LOT hypotheses—including Representational, Computational, Classical, and Nonclassical LOT hypotheses—in terms of the computational architecture they require. This clarifies the empirical commitments of different LOT hypotheses. In Section 3, I rebut some popular arguments for Classical LOT, including the argument that neuroscientific evidence is merely relevant to how computations are implemented. I argue that, on the contrary, neuroscientific evidence is crucial to identifying the correct computational architecture, which in turn is crucial to identifying the correct representations and algorithms. In Section 4, I discuss the degree to which different LOT hypotheses are supported or undermined by neuroscientific evidence about computational architecture. I argue that Classical LOT (Fodor 1975) is ruled out. The only empirically supported version of LOT is the Nonclassical LOT thesis that human brains are capable of cognitive processes that support and are supported by natural language, and hence some neural representations involved in such processes mirror some of the structure of natural language and represent in a language-like way, yet they encode information nondigitally and are processed by ordinary (nondigital, and hence Nonclassical) neural computations that rely not only on syntactic structure but many other features. I wrap up in Section 5. Please note that providing a detailed model of a Nonclassical LOT goes beyond the scope of this report, although I will refer to relevant scientific work when appropriate. The aims of this report are to expand the conversation so that Nonclassical LOT is recognized, clarify the role of computational architecture, and defend the eliminative induction against Classical LOT in favor of Nonclassical LOT.

I hasten to add that Classical LOT has been criticized before, and many past critics have pointed out that it clashes with neuroscientific evidence (e.g., Dennett 1978; Churchland 1992; Horgan and Tienson 1996; Bechtel and Abrahamsen 2002; Matthews 2010; De Brigard 2015). Nevertheless, as we shall see, Classical LOT has continued to be defended and the matter is far from settled.[6] In addition, the recent literature is often unclear about what Classical LOT is committed to, what a Nonclassical LOT amounts to (cf. Aydede 1997, fn. 51), or how LOT hypotheses should be tested. I will sharpen these questions by placing Classical and Nonclassical LOT in the broader context of Representational and Computational LOT simpliciter and by relying on recent advances in our understanding of computation and representation, including the recognition that there are many types of nondigital, and hence Nonclassical, computation

I use the term, is similar to what Pylyshyn (1984) calls "functional architecture". It should not be confused with what Classicists sometimes call "cognitive architecture," by which they tend to mean the symbols and basic computational operations posited by their theory (cf. Fodor and Pylyshyn 1988, 10). This so-called "cognitive architecture" is still at Marr's algorithmic level.

[6] Part of the reason might be that some of the most popular critiques are dubious. For instance, many have argued that brains aren't digital computers because digital computers are serial while brains are "massively" parallel. But the ways in which a computing system is serial versus parallel is a delicate matter that is relative to level of organization. In fact, conventional digital computers can exhibit several forms of parallelism, including "massive" parallelism at the circuit level, which is the level most directly comparable to neural networks (cf. Piccinini 2015, Sect. 13.5).

(e.g., Piccinini 2015, Shagrir 2022). As a result, we will see more clearly why and how neuroscientific evidence refutes Classical LOT but supports Nonclassical LOT. Or so I hope.

## 2. Varieties of LOT and their Representation and Architecture Requirements

A LOT hypothesis can have broader or narrower scope depending on how much cognition it applies to. It could apply to all cognition, many cognitive processes, or only a few. It could apply only to humans or also to other species. For instance, a LOT hypothesis restricted to human internal vocalizations that, introspectively, feel like linguistic episodes is plausible but also relatively narrow in scope. In fact, I will conclude that something close to this is the only empirically well-supported LOT hypothesis. In contrast, a LOT hypothesis that applies to many cognitive processes in many species including insects (e.g., Gallistel and King 2009) is much more ambitious and correspondingly harder to establish.

For our purposes, the important versions of LOT are the following:

> **Representational LOT**: Some cognitive states represent their targets in a language-like way (e.g., Ockham 1323).[7]
>
> > **Computational LOT**: Representational LOT + LOT is processed computationally (Sellars 1954, 1956, 1960, 1962).[8]

---

[7] Many other historical figures besides Ockham defended Representational LOT. Ockham's defense is probably the best known in the contemporary literature, partly due to his discussion by Geach (1957, 101-6). Panaccio (2017) surveys the ancient and medieval literature.

[8] Sellars defends Computational LOT as a useful analogy:

> [The] learning of a language or conceptual frame involves the following logically (but not chronologically) distinguishable phases:
> (a) the acquisition of S[timulus]-R[esponse] connections pertaining to the arranging of sounds and visual marks into patterns and sequences of patterns. (The acquisition of these "habits" can be compared to the setting up of that part of the wiring of a calculating machine which takes over once the "problem" and the relevant "information" have been punched in.)
> (b) The acquisition of thing-word connections. (This can be compared to the setting up of that part of the wiring of the machine which enables the punching in of "information.") (Sellars 1954, 333)
>
> [T]he theory is to the effect that overt verbal behaviour is the culmination of a process which begins with 'inner speech' [i.e., thoughts] … there are many who would say that it is already reasonable to suppose that these thoughts are to be 'identified' with complex events in the cerebral cortex functioning along the lines of a calculating machine. (Sellars 1956, 186-8)
>
> Suppose such an anthropoid robot to be 'wired' in such a way that it emits high frequency radiation which is reflected back in ways which project the structure of its environment (and its 'body'). Suppose that it responds to different patterns of returning radiation by printing such 'sentences' as 'Triangular object at place p, time t' on a tape which it is able to play over and over and to scan.[omitted footnote] Suppose that, again by virtue of its wiring diagram, it makes calculational moves from 'sentences' or sets of 'sentences' to other 'sentences' in accordance with logical and mathematical procedures (and some system of priorities) and that it prints these 'sentences' on the tape. (Sellars 1960, 51-2)

> **Classical LOT**: Computational LOT + LOT representations and computations are digital (Harman 1968a, 1973; Fodor 1968, 1972, 1975; Pribram 1971; Fodor, Bever, and Garrett 1974).
> **Nonclassical LOT**: Computational LOT + LOT representations and computations are *non*digital.

Representational LOT maintains that cognition involves language-like representations. There are two broad classes of relevant languages: (i) natural languages and (ii) formal languages from logic and computer science. Different versions of LOT rely on analogies between thought and either natural languages or formal languages, including the sort of machine language that runs on program-controlled digital computers.

The original LOT hypothesis draws an analogy between thought and natural language (Ockham 1323; Sellars 1956, 1960, 1968; Harman 1968a, 1970a, 1973).[9] Like linguistic utterances, mentalese structures might be made of words that can play the roles of subjects, predicates, etc. and can combine into structures that can play the roles of atomic sentences, which in turn can combine into something equivalent to complex sentences via something equivalent to logical connectives. Like linguistic utterances, mentalese structures might have a compositional semantics and inferential roles that facilitate inference, whereby mentalese conclusions can be generated from mentalese premises in ways that are either deductively valid or inductively justifiable.

Based on such analogies, some philosophers influenced by Sellars argue that acquiring natural language is what allows human beings to engage in propositional thinking (e.g., Brandom 1994; McDowell 1994; Gauker 2011). A related and influential view is that mentalese is not just *analogous* to natural language but *just is* natural language that has been learned, internalized, and is used as a vehicle of thought (Vygotsky 1934/2012).[10] Vygotsky's followers (like Sellars's) typically eschew the "LOT" label, probably because, after Fodor (1975) embraced it, it has become associated most closely with his version of Classical LOT. But Vygotskians' idea that natural language, once internalized, acts as a cognitive scaffolding or neuroenhancement (e.g., Rumelhart et al. 1986; Dennett 1991; Clark 1998, 2008; Lupyan and Bergen 2016; Tomasello 2019; Dove 2020; Borghi 2023; Kompa 2024a) is a version of mentalese *as natural language* (cf.

---

> But to point to the analogy between conceptual thinking and overt speech is only part of the story, for of equally decisive importance is the analogy between speech and what sophisticated computers can do, and finally, between computer circuits and conceivable patterns of neurophysiological organization. (Sellars 1962, 33) also Kaye 1995; Munroe 2025). Another version of LOT as natural language is that mentalese is an internal language ("I-language") made possible by an innate language faculty unique to humans, which allows humans to acquire and process natural language ("E-language"; Chomsky 1986, 1993; on an innate language faculty, cf. Chomsky 1957, 1965). In contrast, the best-known LOT hypothesis is that mentalese is *distinct* from natural language, needed to acquire and understand natural language (Katz 1966), and analogous to the formal languages of logic and computer science (Fodor 1968, 1972, 1975; Pribram 1971; Fodor et al. 1974).[11] This last view comes with at least one additional possible (non-mandatory) analogy: if mentalese is like digital computers' programming languages, mentalese *programs* might control computations in the way that ordinary computer programs control computations.

[9] At least in the philosophical literature, Sellars (1956) marks the transition from more traditional versions of Representational LOT, based on introspection or armchair reflection, to LOT as a scientific of quasi-scientific model aimed at explaining cognition.

[10] Despite the similarities between Sellars and Vygotsky's views, Sellars appears to have developed his ideas about mentalese without knowing about Vygotsky's, perhaps because Vygotsky's main work (1934) was not published in English until 1962. Both Vygotsky and Sellars knew of Dewey (Carl Sachs, pers. comm.), who already argued that "psychic events, such as are anything more than reactions of a creature susceptible to pain and diffuse comfort, have language for one of their conditions" (Dewey 1925, 169).

Both natural and formal language inspirations for LOT share a common core. To be language-like, a system of representations must have, at a minimum, the following features: constituents that play the semantic roles of subjects and predicates within a sentence; when such constituents are combined correctly, they form structures that play the semantic roles of atomic sentences; and when atomic sentences are combined correctly, they form structures that play the semantic roles of complex sentences. Of course, ordinary languages have more structure than that; for example, they have quantificational operators such as "all" and "some", which augment their expressive power; so, language-like representations may have a richer semantic structure than the minimal structure I articulated. And, in addition to their semantic (representational) properties, language-like representations may have syntactic structure and inferential roles analogous to those of linguistic structures.

The debate about LOT is tied to the debate between empiricism and nativism. People on Sellars and Vygotsky's side often lean towards the empiricist view that natural language is acquired via domain-general learning. In contrast, people on Chomsky and Fodor's side often lean towards the nativist idea that natural language is acquired via innate, language-specific processes. Fodor's nativism is somewhat mandated by his reliance on analogies between mentalese and computer code, because it's implausible that cognitive systems could come to think via something like a computer code by *learning* it in the absence of any such computer code in their environment. Other associations are optional. Specifically, we should avoid the misconception that, if thought is like natural language, acquiring mentalese requires acquiring natural language first. That is one possible view, yet it's not mandatory. Even when the analogy is with natural language, mentalese may well be at least partially innate—some cognitive systems may contain an innate mentalese that allows them to *have* propositional thoughts even prior to and somewhat independently of *acquiring* the ability to process a (public) natural language, and possessing mentalese may even be a necessary condition for acquiring (public) natural language. Nevertheless, such an innate mentalese may be analogous to natural language. Roughly, that is Chomsky's view. Of course, even if mentalese is both partially innate and either analogous to natural language or the same as the neural representations involved in acquiring natural language, it may well be that, as Vygotsky's followers argue, acquiring the ability to use a (public) natural language augments our thinking prowess. At any rate, *the extent to which mentalese is innate is a difficult empirical question on which I remain neutral*.

[11] Both Fodor and Chomsky maintain that mentalese is supported by in an innate language faculty, but Chomsky denies that mentalese is distinct from natural language: "[i]t is often argued that another independent language of thought [i.e., independent of natural language] must be postulated, but the arguments for that do not seem to be compelling" (Chomsky 2007, 16; cited by Dupre 2021, 774; cf. Hinzen 2013). Fodor, Bever, and Garrett attribute to Pribram (1971) the view that "there is a language 'of thought' (or 'of the neurons') which is different from the language we speak, and that speaking involves the encoding of messages which are originally formulated in that language" (Fodor et al. 1974, 376). Pribram does defend the idea that humans think by means of a language analogous to computer programming languages (Pribram 1971, esp. Ch. 19). Harman attributes to Katz (1966) the view that "when a person speaks, he encodes his thoughts in his language; and when he understands someone else, he must decode what the other has said by translating it into the basic language of thought" (Harman 1968b, 20). Harman's interpretation seems fair even though it goes somewhat beyond what Katz says.

Representational LOT, by itself, says nothing more about the properties of mentalese and offers no mechanism for how cognitive states are processed in accordance with their language-like structure and semantic content. In this respect, Representational LOT is compatible with whatever neuroscience finds in the brain provided that, in the relevant cases, there are language-like neural representations. Thus, Representational LOT per se has no architectural requirements because it's not yet a computational hypothesis; all it requires is that some neural representations be language-like.[12]

Computational LOT adds to Representational LOT a generic computational requirement, according to which mentalese is processed computationally. While what counts as physical computation is controversial (Anderson and Piccinini 2024), three aspects of computation are relatively uncontroversial and sufficient for our purposes. First, computations are physical processes that can manipulate representations in accordance with their semantic content. Second, there are different kinds of computation—digital, analog, neural, etc.—that involve different kinds of representations and algorithms. Third, different kinds of representations and algorithms require different computational architectures. Examples include the so-called von Neumann architecture (von Neumann 1945; Patterson and Hennessy 2011), which can process digital representations in accordance with suitable digital algorithms encoded as programs and stored in long-term memory, and the general-purpose analog computer (Pour-El 1974), which can process analog representations in accordance with suitable analog methods. By positing that mentalese representations are processed by an appropriate computational architecture, Computational LOT constrains the physical mechanism for processing mentalese—Computational LOT requires computational machinery that can combine language-like representations and manipulate them in the requisite way.

Computational LOT is committed to there being language-like representations in the brain as well as components capable of computing over them in accordance with their semantic properties.[13] Computational LOT simpliciter simply asserts that some form of computation is enough to process mentalese in the right way (Sellars 1954, 1956, 1960, 1962).[14] Beyond this rather vague requirement, the exact features of the representations and hardware posited by Computational LOT can be left up to neuroscience to discover. Crucially, Computational LOT simpliciter does not require that primitive computational operations be defined over the semantically primitive LOT constituents (i.e., mentalese words).[15] All that matters for Computational LOT to be true is that the LOT constituents be encoded and processed in ways that accord with their semantic content. They could be encoded in layers of ordinary neural representations and processed via ordinary neural computations (more on this in Sect. 4).

[12] For an attempt to develop Representational LOT in a *non*computational direction, see Horgan and Tienson 1996.

[13] Strictly speaking, there are LOT hypotheses that attribute language-like structure to thoughts, possibly in combination with suitable computational processes, while *rejecting* Representational LOT (e.g., Stich 1983). In this essay, I consider only LOT hypotheses that entail Representational LOT.

Classical LOT adds a digitality requirement that allows LOT constituents—particularly, mentalese words—to be encoded as semantically atomic yet digitally encoded "symbols" and processed via computational operations defined over the syntactic properties of the symbols (Fodor 1975; Newell and Simon 1976; Pylyshyn 1984). The term "symbol" is highly ambiguous and, regrettably, it is rarely disambiguated in the LOT literature. It has at least four relevant meanings: (i) *representation* simpliciter, (ii) *representation with an arbitrary semantic content* (this is Peirce's notion of symbol, which contrasts with icons, whose content is due to something like resemblance, and indices, whose content is due to something like causation; Atkin 2023), (iii) *amodal representation*, and (iv) *digitally encoded representation*. Let me briefly explain why the fourth meaning is the most relevant to assessing Classical LOT.

It is experimentally well established that neurocognitive systems rely on representations (e.g., Thomson and Piccinini 2018). Since human languages and other communication systems are symbolic in Peirce's sense, presumably the neural representations involved in processing language and other symbolic communication systems are symbolic in that sense (more on this in Sect. 4.1). Surely there is more to Classical LOT than this basic point. An amodal representation is one that is not tied to any specific sensory modality (Wajnerman Paz 2017); whether a representation is amodal is orthogonal to whether it is language-like (cf. Calzavarini 2025). That is, natural language itself can be represented within various sensory modalities including auditory (spoken language), visual (written language), or tactile (Braille), while neural representations may be amodal without being language-like (e.g., Tamber-Rosenau et al. 2013) and the modal-amodal distinction might even be an idealization that comes in degrees (Michel 2021). Thus, that some neural representations are Peircean symbols is insufficient for Classical LOT, and whether they are amodal is orthogonal to Classical LOT.

Classical LOT requires something more specific: it requires *digitally encoded* representations of the sort found in formal logic and classical computability theory—the kind that can be processed via digital computation—or else it collapses into generic Computational LOT. As we've seen, Classical LOT draws an analogy with such formal languages. The main defining feature of formal languages—what allows them to be processed via digital computation—is that their structures are composed of digitally encoded Peircean symbols. Accordingly, from now on, when I refer to the symbols posited by the Classical LOT hypothesis, I mean digitally encoded Peircean symbols.

[14] In fairness, Sellars's analogy appears to be between thought and language-like *digital* computing. Crucially, it is explicitly a partial *analogy*, not a claim that cognitive processes are literally digital computations. Therefore, his view leaves room for the possibility that the analogy holds between thought and *computation* simpliciter, and hence that thoughts and their neural realizers be nondigital.
[15] I use "words" to denote these semantically primitive mentalese constituents regardless of whether they would map onto words in natural language.

Therefore, in addition to the assumptions that thoughts represent in a language-like way (Representational LOT) and are processed computationally (Computational LOT), Classical LOT adds that the computational architecture that encodes and processes mentalese structures is like a digital computer. This, in turn, implies that, like the digital representations processed by digital computers (Piccinini 2015), mentalese structures are concatenations of finitely many types of digits. By "digit", I mean a discrete state that belongs to one of finitely many types that the system can distinguish reliably from other types (digits do not come in degrees), and which can be concatenated with other digits so that it's clear which digit comes first, which is next, and so forth until the last digit in any composite representation.

It's important to appreciate two things about digital encoding. First, digital encoding of language-like representations is such a core commitment of Classical LOT hypotheses that it's usually left implicit. Yet it's a nonnegotiable commitment. In fact, Classical LOT theorists typically discuss computation as if digital encoding were a necessary but not even sufficient condition for computation, and thus nothing can be computational without being digitally encoded.[16]

Second, digital encoding is a demanding requirement, which goes beyond encoding information into *discrete* representations. Cognition is uncontroversially *categorical*. That is, humans and other animals often organize information that may vary along a continuum into discrete categories (e.g., Cesanek et al. 2023). On the sensory side, cognizers often experience a sharp shift in perception at certain points along a physically continuous spectrum, rather than a gradual change. On the action side, cognizers produce categorically distinct responses (e.g., walking, jogging, running) rather than responses that vary along a smooth continuum. To engender categorical perception, neurocognitive systems amplify differences near category boundaries, perhaps through a combination of lateral inhibition, top-down modulation, and representational clustering, even though neural representations themselves are typically graded (and hence nondigital; more on this in Sect. 4.3). To engender categorical motor control, neurocognitive systems rely on different neural circuits for different action patterns and possibly global inhibition (cf. Penconek 2025). Even more generally, neural systems often encode information in the activity of neural assemblies (Yuste et al. 2024)—group of neurons that fire together and trigger other assemblies to form complex patterns of activity.[17] Therefore, a representational system can separate targets that vary along a continuum into discrete categories without being language-like and without encoding information digitally (cf. Block 2023, esp. Ch. 6, to which I am indebted here). I submit that categorical representation is enough for neurocognitive systems to represent and process mentalese, without needing to encode its syntax or other properties digitally let alone process it via digital computation. Meanwhile, the Classical LOT hypothesis maintains just the opposite: mentalese is represented and processed by digitally encoding its syntax.

---

[16] For instance, Fodor agreed that neural processes appear to be nondigital and concluded that, therefore, they are not computational: "what is usually characterized as computational neurology isn't computational" (Jerry Fodor, pers. corr., 2005). As I will point out shortly, the kind of narrow conception of computation to be found in much of Fodor's work is no longer tenable (Piccinini 2015, Shagrir 2022).

[17] Such patterns are sometimes called "neural syntax" (e.g., Buzsáki 2019), which should not be confused with the kind of linguistic syntax at issue here.

Digital encoding is neither necessary nor sufficient for language-like representation. It's insufficient because we can digitally encode representations that are not language-like (e.g., Johnson-Laird 1983). It's unnecessary because language itself can be encoded via nondigital representational systems such as cursive, pictographs, logographs, or ideographs. That said, language-like systems *can* be encoded digitally. Therefore, *if* the right hardware and software were present within brains, Classical LOT *would* explain how cognitive computations process language-like structures in accordance with their content: by performing primitive digital operations on mentalese words and sentences based on their (digitally encoded) syntactic structure. For, as Turing (1936-7) and other logicians showed, digital computations can be defined over (digital encodings of) syntactically structured representations in such a way that certain semantic relations (e.g., of entailment) between such representations are respected. Fodor and collaborators are especially insistent on this point and often *define* LOT simpliciter as a model that posits computations sensitive to the combinatorial syntax of mental representations, as if "Nonclassical LOT" were an oxymoron.[18] Digital computers dominate computing technology and digital computation used to dominate cognitive science, or at least Classical cognitive science. Accordingly, at least historically, most defenders of LOT restricted their attention to Classical LOT.[19]

---

[18] Here is an example:

> Classical theories—but not Connectionist theories—postulate a 'language of thought' (see, for example, Fodor, 1975); they take mental representations have a *combinatorial syntax and semantics*, in which … the semantic content of a (molecular) representation is a function of the semantic contents of its syntactic parts, together with its constituent structure… In Classical models, the principles by which mental states are transformed, or by which an input selects the corresponding output, are defined over structural properties of mental representations (Fodor and Pylyshyn 1988, 12-13; emphasis original).

Here is another way they put the point:

> It would not be unreasonable to describe Classical Cognitive Science as an extended attempt to apply the methods of proof theory to the modeling of thought (and similarly, of whatever other mental processes are plausibly viewed as involving inferences; preeminently learning and perception) (Fodor and Pylyshyn 1988, 20-21).

This is meant to map to something in the brain:

> [T]he symbol structures in a Classical model are assumed to correspond to real physical structures in the brain and the combinatorial structure of a representation is supposed to have a counterpart in structural relations among physical properties of the brain (Fodor and Pylyshyn 1988, 13)

[19] Even Harman did so:

To properly assess LOT, we must also include Nonclassical LOT hypotheses, especially since I will argue that they are the only viable ones. For present purposes, to a first approximation, a LOT hypothesis is maximally Nonclassical just in case it posits mentalese representations and computations that are *non*digital. The notion of computation is more general than the notion of *digital* computation. For starters, there is analog computation, which was developed well before modern digital computers. Around the time that McCulloch and Pitts (1943) modeled the brain as a digital computing system, Craik (1943) and others (Gerard 1951; Lashley 1958; cf. von Neumann 1958) suggested that the brain is more like an *analog* computer. In contrast, later I will argue that neural computation is sui generis (per Piccinini and Bahar 2013). For now, the important point is that different kinds of computation require different kinds of computational architectures that manipulate different kinds of vehicles according to different kinds of algorithms.

As we've seen, the analogy between thought and formal languages is primarily due to the appeal of digital computation in the first place. Therefore, if the digitality requirement is dropped, there is hardly any reason to invoke analogies with formal languages, and the most natural analogy becomes between thought and *natural* language. As we've already seen, Sellars defends a LOT hypothesis of the latter sort, which is potentially Nonclassical. Many subsequent authors may also fit into this Nonclassical LOT camp (for a recent proposal, see Wu et al. 2024).

To avoid confusion between different Classical architectural assumptions, let's subdivide Classical LOT into at least three versions of increasing strength:

> **Weak Classical LOT**: Classical LOT + Cognition is carried out by digital computing systems aka automata (McCulloch and Pitts 1943,[20] Harman 1973).

---

> An abstract automaton is specified by its program. The program indicates possible reactions to input, how internal states plus input can yield other internal states, and how internal states and input can lead to various sorts of output (Harman 1973, 42).
>
> We were led to see a person as an automaton. To understand a type of mental state or process is to see what function such states or processes can have in a person's "program." … I will speak of a "language of thought" and will speculate on the relations between the inner language of thought and the outer language we speak (Harman 1973, 53-4).

In context, it's clear that by "automaton", Harman means something like a finite state automaton or a Turing machine—a type of digital computing system (cf. Harman 1968a, 594-5).

[20] McCulloch and Pitts (1943) argue that nervous systems are digital computing systems that process propositional representations. In their model, neurons process propositional representations by implementing logical inferences, a view that comes close to Weak Classical LOT, though they didn't talk about encoding or processing syntactic structure. Still, the connection they drew between formal logic, computation, and brain theory is the germ of Classical LOT.

**Moderate Classical LOT**: Classical LOT + Cognition is carried out by special-purpose digital processors distinct from digital memory (Fodor 1972, 1983; Cummins 1983[21]).
**Strong Classical LOT**: Classical LOT + Cognition is carried out by program-controlled digital processors (Miller, Galanter, and Pribram 1960; Fodor 1968, 1975; Fodor, Bever, and Garrett 1974; Newell and Simon 1976; Pylyshyn 1984).

To understand the hardware requirements of these three Classical LOTs, we need to distinguish three relevant kinds of digital computing systems: hardwired, plastic, and program controlled. Hardwired digital computing circuits, like those found in ordinary computer processors, perform fixed operations. Typically, changing the operations performed by such circuits requires physically rewiring the components. This is what programming a computer consisted of in some early computers: manually plugging cables to connect components in the desired way. Of course, in the brain there are no programmers to manually rewire circuits. And we know that cognition requires flexibility: different operations need to be selected depending on task, evidence, and so forth. We also know that neural circuits are plastic—they change their input-output function over time. Therefore, setting aside the nonactual possibility that brains have a fixed structure, there are two main options consistent with Classical LOT: either neural circuits are plastic—that is, they can change their own organization over time—or they divide into processing and control components that constitute a (collection of) program-controlled, digital processor(s).

The former view, according to which the relevant circuits are plastic and self-organize, is more closely associated with Connectionism and computational neuroscience, but it may also be combined with Weak or Moderate Classical LOT (more on this in Sect. 4).[22] In contrast, the latter view, according to which cognitive processes are executions of programs stored in memory banks separate from the processors, is Strong Classical LOT (Fodor 1968, 1975; Newell and Simon 1976; Pylyshyn 1984).

We can now sketch the representation and architectural requirements of Classical LOT. All versions of Classical LOT share the same representation requirement: a system of digital representations that encode language-like syntactic structures processed via primitive digital operations defined over semantically atomic symbols. This, in turn, requires a finite list of words (atomic symbols) plus rules for combining the words into atomic sentences and atomic sentences into complex sentences. Typically, digitally encoded language-like systems build words out of letters from a finite alphabet, though the letters carry no semantic information by themselves. The use of an alphabet increases the efficiency of the system but is not strictly necessary.[23] This representation requirement leads to Weak Classical LOT's architectural requirement: a system of digital components whose states can encode the mentalese constituents (words, sentences)—including the concatenation between them that realizes their combinations—and can process those representations. I will argue that these general requirements are not satisfied by biological brains.

[21] Cummins (1983) accounts for computation in terms of program execution but then accounts for program execution in terms of merely acting in accordance with a program, so his view sounds like Strong LOT but is probably a version of Weak or Moderate LOT.

[22] Classical LOT is often contrasted with Connectionism, computational neuroscience, or both. By "Connectionism", I mean a framework that uses artificial neural networks to provide how-possibly explanations of cognitive capacities without being constrained by neuroscientific evidence. By "computational neuroscience", I mean a framework that appeals to biological neural structures (including but not limited to neural networks) to explain cognitive capacities and is constrained by neuroscientific evidence. Thus, as I define them, in principle both Connectionism and computational neuroscience are compatible with Classical LOT, and neither of them is committed to cognition being solely a matter of association, although typical Connectionist and neurocomputational models are Nonclassical.

Weak Classical LOT has no provision for the separation of processors and memory. The separation of processors and memory is sometimes implicit and sometimes explicit in formulations of Moderate Classical LOT. Let's make it explicit. Moderate Classical LOT distinguishes processors and memory, which stores representations for extended periods. Coordinating between digital processors and memory requires sophisticated control components that keep track of what is stored in different memory locations, which operations need to be performed on data structures and in which order, and which data structures need to be either fetched from or stored within specific memory locations at any given time.

In contrast to Weak and Moderate Classical LOT, Strong Classical LOT posits full-blown general-purpose processors that execute *programs* stored in memory on data structures also stored in memory. From a distance, the architectural requirements of Moderate and Strong Classical LOT look similar. In fact, Strong Classical LOT requires less hardwiring of the control structure because much of the control is delegated to instructions that can be stored in memory in the same form as the data.

Having outlined the distinction between Classical and Nonclassical LOT and their architectural requirements, we now turn to the arguments and evidence for and against LOT. The next step is to tackle some general arguments that have been given for Classical LOT.

## 3. Architectural Arguments and The Implementation Objection

There are some general arguments for Classical LOT based on architectural features that are allegedly needed to explain cognition. Most notably, Minsky and Papert (1969) proved that two-layer perceptrons cannot compute relatively simple 2-bit Boolean functions such as XOR, Fodor and Pylyshyn (1988) argued that certain associative neural networks cannot account for the alleged productivity and systematicity of cognition, and Gallistel and King (2009) argued that synaptic strength cannot account for animal memory.[24] Since Classical LOT can be used to compute XOR and explain productivity, systematicity, and memory, Classical LOT is sometimes thought to follow. I call these considerations *architectural arguments* for Classical LOT.

---

[23] In McCulloch and Pitts's 1943 model, there aren't even words. Each atomic symbol encodes a fully propositional content. That is not yet a true LOT hypothesis as I understand it here; sub-sentential structure is indispensable.

[24] Of these, systematicity is probably the most discussed (e.g., Aizawa 2003; Calvo and Symons 2014). For recent evidence that Nonclassical architectures can exhibit the requisite degree of systematicity, see O'Reilly et al. 2022; Lake and Baroni 2023; von der Malsburg 2024. For an excellent critique of Gallistel and King's argument, see Morgan 2019.

Note that architectural arguments do not provide observational evidence that brains contain the representations and architecture posited by Classical LOT; they only provide an inference to what is alleged to be the best (or only) explanation. Their structure is the following:

(P1) Classical LOT can explain cognitive capacity C;
(P2) Nonclassical architecture M cannot explain C;
(C) Therefore, Classical LOT is the likely explanation of C.

The original proponents of architectural arguments may or may not have intended them to *prove* that Classical LOT is the *only* possible explanation of cognition. Since at least some of them are often interpreted as such, or at least as providing strong evidence for Classical LOT, we need to address them as such. If architectural arguments succeed, then the debate over LOT is settled regardless of how brains turn out. Let's see why they fail.

Luckily, we do not need to get into the details of what each architectural argument purports to show or whether any of the specific alternatives they consider can or cannot explain the phenomena in question. The reason architectural arguments fail is that *even if* M cannot explain C, Nonclassical architectures other than M might.

The failure of architectural argument can be boiled down to the following computability considerations. We know that brains can compute functions such as 2-bit XOR. We know they can store information and expand their use of memory resources up to a point. At least some nervous systems show behaviors that exhibit some degree of productivity and systematicity. These are important, coarse-grained facts about neural computation that can be inferred from behavior alone, or perhaps from behavior in combination with general theoretical considerations. Such facts constrain our understanding of neurocomputational architecture and virtually no one disputes them.

Nevertheless, brains are finite. Therefore, when unaided by external memory storage, at any given time, brains are computationally equivalent to (very large) finite state automata (FSA). What I mean is that, whether or not brains use digital representations and computations (as Classical LOT alleges), insofar as their computational capacities can be *modeled* by computability theory, their computational capacity can be modeled by FSA (though see Maley and Piccinini 2016 for some subtleties). This constrains what the computational architecture of nervous systems is capable of, but it determines neither the format of neural representations nor other details of neurocomputational architecture.

In general, the same outer behavior is compatible with many inner structures. In other words, inner structure is underdetermined by outer behavior. A theory of the brain's computational architecture is no different. The mind sciences, especially when they appeal solely to behavioral benchmarks without looking under the hood, are especially prone to underdetermination (cf.

Anderson 1978 for a classic case study). The Classical LOT hypothesis is a case in point—a cautionary tale that illustrates how risky it is to speculate about internal structure solely on the basis of behavioral evidence and theoretical considerations, or relatively successful models, without considering available evidence about inner structure.

Most relevantly, there are lots of ways to build FSA-equivalent devices, some digital and program-controlled, some digital and hardwired, some digital and plastic, many nondigital, some known, and many unknown. Even if a specific type of FSA-equivalent system (say, one that must use only two-layer perceptrons, certain types of associative neural networks, or synaptic strength as an account of memory) fails to explain some cognitive capacity, it doesn't follow that another specific type of FSA-equivalent system (say, one that uses a Classical LOT) is the way the brain works. Therefore, general architectural arguments for Classical LOT do not establish that Classical LOT is the correct explanation of cognition, or even a plausible one. The computational architecture of the brain cannot be discovered via general considerations merely based on behavioral benchmarks; it can only be discovered by examining brains and figuring out which computational architecture is there.

Here, defenders of Classical LOT might reply that I am missing the point of the architectural arguments. They might suggest that no matter what the performance limits of nervous systems might be in practice, a theory of cognition investigates the *competence* of the system. In turn, architectural arguments strongly suggest that the competence of cognitive systems is computationally equivalent to a universal Turing machine, and universal Turing machines process digitally encoded data and programs. Any system with the right competence must be computationally more powerful than any FSA and, therefore, it must process a Classical LOT.[25]

This is still a non sequitur, and my response is three-fold. First, not every Classical LOT hypothesis posits a computationally universal architecture; only Strong Classical LOT does.[26] And Strong Classical LOT is much less plausible than Moderate Classical LOT (more on this in Sect. 4.3). Thus, the more plausible version of Classical LOT—Moderate Classical LOT—is inconsistent with the above reply. Second, while architectural considerations are hugely important (more on that in Sect. 4), here we are just focusing on sheer computational capacity. Sheer computational capacity, or any more specialized capacity considerations such as the need to explain productivity and systematicity to the extent that cognizers exhibit it (cf. Johnson 2004), does not support Classical LOT.

---

[25] Some authors have speculated that brains might be computationally *more* powerful Turing machines, which of course are computationally more powerful than FSAs (e.g., Copeland 2000). But no mechanism by which such hypercomputational power could be achieved has been proposed, and no evidence that cognizers have hypercomputational powers has been given. Whether cognition is hypercomputational should not be confused with whether cognition includes aspects that are Turing-uncomputable, e.g., because they include random aspects (cf. Turing 1950; Piccinini 2003, 2020a).

[26] Strictly speaking, even Strong Classical LOT, as I defined it, need not posit a computationally *universal* architecture; it only needs an architecture that is program-controlled. A program-controlled architecture may or may not be universal, depending on whether it supports a universal set of instructions. This does not affect the thrust of my argument, so I set it aside.

Third, even if there were a sense in which the competence of (idealized) cognitive systems is computationally universal, Classical LOT in the most important sense would still not follow. Computational universality depends on three architectural properties: unbounded memory, the ability to store data and instructions, and the ability to control computations in response to instructions. None of this requires the core Classical assumptions that instructions and data be encoded and processed digitally, and that memory and processing functions be carried out by structurally separate components.

In fact, there *is* a clear and uncontroversial sense in which idealized, linguistically competent humans *are*, indeed, computationally universal, though in a way that is limited in practice by their small memory capacity. Humans can follow any sequence of linguistic instructions they can memorize on any data they can memorize, thereby being equivalent to computationally universal systems until they run out of memory (which is usually pretty quickly). For all we know, humans may accomplish this by encoding and processing data and instructions nondigitally within neural systems that fulfill both memory and processing functions. In other words, insofar as human neurocognitive systems are computationally universal, they might accomplish this feat by means of a Nonclassical computational architecture. At any rate, cognitive systems are still limited by their finiteness, and hence equivalent in practice to an FSA—unless aided by an unbounded external memory, which is beside the point.

In addition, later we'll see that computational architectures that have little or nothing to do with Classical LOT are better than Classical LOT architectures at processing human language itself—the very capacity that inspired LOT in the first place. To find out how neurocognitive systems exhibit their competence, we must investigate their actual computational architecture, including but not limited to the format of their representations, the type of processors they use, the operations they perform, the type of memory they have, and whether memory is separate from processors. And the only way to discover our neurocomputational architecture is to study how brains work.

There remains what I call the *implementation objection*. This is the idea that the best that "Connectionist" (and, presumably, neurocomputational) theories can hope for is to explain how the algorithms and representations posited by Classical LOT are implemented (Fodor and Pylyshyn 1988, pp. 64-6). The implementation objection goes hand in hand with the architectural arguments. We've already seen that the architectural arguments carry no water. Let's briefly diagnose what goes wrong with the implementation objection.

The implementation objection assumes that all the evidence pertaining to a theory of cognition belongs either at the algorithmic level (algorithms and representations) or the implementation level. Furthermore, the objection assumes that "psychology" (i.e., behavioral evidence) is solely in charge of the algorithmic level.[27] Ergo, any evidence that is supposedly not about algorithms and representations—i.e., all the evidence from neuroscience—must be about implementation. This misconstrues the situation in at least two crucial ways. First, as I pointed out, there is a level of analysis between algorithms and implementation: computational architecture (e.g., processors, memory, and how they are organized). Second, the hardware constrains the architecture, which in turn constrains the algorithms and representations it can run.

[27] Cf.: "[I]n the language of neurology…, presumably, notions like computational state and representation aren't accessible" (Fodor 1998, 96). Of course, nothing could be further from the truth.

For instance, ordinary digital computers are made of logic gates arranged to constitute processors and memory. Computer engineers call this "logic design". This is the computational architecture of digital computers. One and the same computational architecture can be implemented using vacuum tubes, electromechanical relays, various kinds of integrated circuits, or some other means. The latter are the physical technologies used to build digital computing hardware. If the hardware—whatever it is—lacks appropriate degrees of freedom and organization, it fails to constitute the relevant architecture. Specifically, if a system contains no hardware components that implement logic gates, it cannot be a von Neumann architecture. And without the relevant architecture, digital computations cannot run. Another example of computational architecture is the use of analog integrators and other analog components (adders, multipliers, etc.) to solve differential equations within a general-purpose analog computer. Analog integration can be physically implemented using balls rotating on a disc or operational amplifiers. But if a physical system contains no hardware components that implement analog integration (and other analog operations), the system cannot be a general-purpose analog computer.

To reiterate, computing systems can be analyzed at *four* levels of analysis: the computations they perform, the algorithms and representations they use to perform them, the computational architecture that processes the representations in accordance with the algorithms, and the technology that implements the architecture. Just because a computational architecture can be implemented using different technologies, it would be fallacious to conclude that any physical technology, no matter how arranged, constitutes a computational architecture capable of running digital representations and algorithms.

Levels of analysis constrain one another. In one direction, algorithms and representation of any given type can only run on architectures capable of running them, and a given architecture can only be implemented on hardware with relevant properties. In the other direction, if hardware has a certain structure and organization, this constrains the computational architectures it can implement, which in turn can only support certain types of representations and algorithms. Thus, if neural structure and organization do not implement a Classical architecture, the brain cannot run Classical LOT representations and algorithms, and the Classical LOT hypothesis fails. The crucial point is that if neural hardware supporting a Classical architecture is missing, then Classical LOT is false.

Another way to put this point is that "implementation" is ambiguous between implementing algorithms and representations within a computational architecture and implementing a computational architecture using a certain physical substrate. The implementation objection is right only insofar as Classical LOT requires a Classical computational architecture. But whether a

Classical architecture is present in neural systems can only be established by examining neural systems.[28]

In conclusion, both the architectural arguments and the implementation objection provide no evidence whatsoever that Classical LOT is the correct explanation of cognition. At best they establish that, if Classical LOT is true, then brains must possess the relevant computational architecture. Our question, then, is whether the kinds of representation and algorithm posited by any version of LOT, including Classical LOT, and the kinds of computational architecture such representations and algorithms require, are found in the brain.

---

[28] Two anonymous readers objected almost identically:

> [Reader 1] But there is a third option: a physical substrate implements one computational architecture (the kind of information processing observed in cognitive neuroscience) which then implements another computational architecture. We see this all the time in computers, where one computational architecture implements another (and perhaps more—there is no reason why this could only iterate once), and all these hierarchically implemented architectures are realized in the same piece of hardware. This was the claim of Fodor and Pylyshyn: connectionist networks, which are not physical substrates but are a computational architecture, can implement another computational architecture that supports LOT, and this is not an implausible claim about how it works in the brain.
>
> [Reader 2] [A] classicist could basically grant everything the author says about the messy, graded, and stochastic nature of neural wetware, then argue that this neural substrate realizes a virtual machine that is classical and digital. The claim here isn't that neurons are logic gates, but that the organized activity of vast populations of neurons implements the functional profile of a system that processes discrete symbols according to rules. This "virtual machine" defense is about a higher level of organization: the continuous dynamics of the lower-level system could be organized such that, at a higher level of description, its states map cleanly onto the discrete states of a Turing machine or classical symbol-processing system.

Compare Fodor and Pylyshyn:

> *[I]mplementation*, and all properties associated with the particular realization of the algorithm that the theorist happens to use in a particular case, is irrelevant to the psychological theory; only the algorithm and the representations on which it operates are intended as a psychological hypothesis. Students are taught the notion of a "virtual machine" and shown that *some* virtual machines *can* learn, forget, get bored, make mistakes and whatever else one likes, providing one has a theory of the origins of each of the empirical phenomena in question (1988, emphasis original).

These remarks invoke a third notion of implementation: implementing a virtual machine by a physical machine. A virtual machine in this sense is a complex arrangement of data and instructions stored in the memory of a digital computer together with the execution of the instructions on the data. Implementing a virtual machine in this sense requires digitally encoding the virtual machine, storing it in memory locations distinct from the processors, possessing a system of digital addresses for accessing the content of memory when needed, keeping track of the memory locations that need to be accessed at each step, and much more—in other words, a program-controlled version of the Classical architecture that I am questioning (technically, a virtual machine can be stored in virtual memory, whose addresses are automatically translated into physical addresses by the operating system and memory management hardware). As I will argue shortly, neurobiological systems lack all the features needed to support the implementation of virtual machines in this sense. Because this objection presupposes precisely the kind of Classical, program-controlled architecture that is under dispute, it unwittingly begs the question.

## 4. Evidence for and against LOT

I will now briefly review some of the evidence for or against the claim that brains operate in the way required by Classical LOT or, indeed, any LOT.

### 4.1 Evidence for Representational LOT

Let's begin with Representational LOT, the relatively modest claim that some (sequences of) cognitive states represent their targets in a language-like way. There are three types of evidence in its favor.

First, there is outer speech. Humans use both natural and formal languages, including the languages of logic and math; therefore, human brains contain machinery capable of language processing. No one disputes this. So, merely pointing to language use is not enough to support the existence of a mentalese. Yet it is both obvious and a well-established finding of contemporary linguistics and philosophy of language that human linguistic utterances have complex syntactic, semantic, and pragmatic structure, much of which is not explicitly encoded in utterances themselves but must be understood by speakers and listeners based on a broader understanding of the language and context in which utterances are made. Given all this, it's hard to see how humans could acquire the ability to understand and produce language without processing neural representations that mirror both the overt and covert structure of utterances and their linguistic context. In addition, humans acquire the ability to construct and comprehend not only simple utterances but full arguments that employ logical connectives. It's hard to see how humans could acquire this ability without there being something in human brains that tracks the structure of complex linguistic constructions and arguments. Whatever neural representations are involved in language processing must be able to disambiguate ambiguous expressions and represent the syntactic, semantic, pragmatic, and inferential properties of language, and more (e.g., phonology). At any rate, it has been empirically demonstrated that human brains bear neural signatures of utterances during production and comprehension.[29] These are neural representations that occur while people process language. Such neural representations mirror at least some of the structure of linguistic utterances (Fedorenko et al. 2024).[30] There is also evidence that even some non-linguistic perceptual representations involved in natural language comprehension acquire some language-like structure (cf. Bocanegra et al. 2022). This is already enough to establish a version of

---

[29] I first learned this from a talk by Patrick Suppes (cf. Suppes et al. 1999).

[30] Since public language can be ambiguous or otherwise underspecified in a way that language comprehension and production are not, mentalese must contain additional structure not found in public language (cf. Kaye 1995, 101; Hinzen 2015; Dupre 2021, Sect. 3). Dupre (2021, Sects. 5-6) argues that mentalese cannot be natural language, by which he means the i-language posited by generative linguists. His reason is that there are sentences that are ungrammatical yet acceptable; therefore, there might be a medium distinct from the i-language in which to express such sentences. Even if Dupre's conclusion were correct (a big if, given that he himself explores several strategies for avoiding it), this is consistent with the present point. The medium posited by Dupre may well be a type of language-like neural representation partly distinct from the neural representations that underlie the i-language.

Representational LOT: at least the neural representations involved in language production and comprehension represent in a language-like way.[31]

Second, there is inner speech. Many humans report experiencing at least some of their thoughts as if they were a kind of inner monologue, expressed within a natural language they understand and either heard or spoken in their head. This is robust introspective evidence that some cognitive events have language-like structure (in natural language). Yet inner speech need not be introspectable. There is evidence that some thought processes involve unconscious linguistic imagery of the sort that humans can sometimes introspect, that human brains bear neural signatures of inner speech (Jack et al. 2019), and that inner speech enhances some cognitive functions (Nedergaard et al. 2023a, b; cf. Carruthers 1998, 2002; Lagland-Hassan and Vicente 2018; Munroe 2023, 2025; Frankfort 2024; Kompa 2024a, b). This is strong evidence that neural representations involved in inner speech represent in a language-like way.

Third, there are linguistic determinants of human thought. As Hinzen et al. (2024) argue persuasively, natural languages encode grammatical properties such as Person (the difference between *I*, *you*, *they*, etc.) and Tense (the difference between *was*, *is*, etc.), among others, that are not known to have non-linguistic counterparts or to exist independently of natural language and are not reducible to other syntactic or semantic properties. It is independently plausible that humans have at least some thoughts that are expressible in natural language and include such grammatical properties (cf. Hinzen 2013)—if nothing else, as the previous paragraph points out, it is introspectively obvious to many of us that we can recite linguistic utterances in our own inner speech. Therefore, at least the neural representations that realize linguistically expressible thoughts represent in a language-like way. Importantly, the grammatical properties in question are properties of natural language, not of any mentalese distinct from natural language.

Further evidence for language-like representations is hard to come by, especially when it comes to the digitally encoded symbols posited by Classical LOT. I will briefly consider two putative examples.

The first example is a defense of LOT offered by Dehaene and collaborators based on a series of recent studies: "humans possess multiple internal languages of thought, akin to computer languages, which encode and compress structures in various domains (mathematics, music, shape…) … [H]umans […] engage a logical, recursive mode of representation akin to a programing language" (Deheane et al. 2022, 751-2). If their analogy with computer programs is interpreted strictly, this is a version of Strong Classical LOT. Here I cannot do justice to the depth of their work. The following brief considerations will suffice to illustrate how difficult it is to find evidence for LOT, particularly Classical LOT.

[31] The present argument from language acquisition and processing to language-like neural representations should not be confused with superficially similar-sounding arguments from language acquisition and processing to "classical, symbolic architectures" (e.g., Dupre 2023, 410). Classicists argue that only a Classical LOT architecture can acquire and process language. The conclusion of this non sequitur is directly refuted by the fact that our best models of language processing are deep neural networks that use continuous (Nonclassical) representations (cf. Chowdhery et al. 2023; Milliere forthcoming).

The evidence offered by Dehaene and collaborators pertains to sequences, such as regular sound sequences like ABAB… or AABBAABB… (Al Roumi et al. 2023), or regular geometric patterns like zig zags and squares (Sablé-Meyer et al. 2021). Such regular sequences can be generated by nested repetitions of primitive operations. School-educated humans can exploit symmetries and repetitions in such sequences to encode them in a compressed form that depends on the order, number, and nesting of the operations. Subjects memorize such sequences, predict how they will continue, and compare them with deviant sequences better than they do with less regular sequences, whose representations cannot be compressed in the same way. Brain recordings using magnetoencephalography suggest that when it comes to regular geometric sequences, subjects' brains encode not only specific locations but also the geometric properties of transitions between locations and the ordinal position of transitions within a sequence (Al Roumi et al. 2021), strongly suggesting that regular sequences are neurally represented in a compressed form that depends on the order, number, and nesting of the operations. Monkeys appear to either lack such capacities for compressed representations of regular sequences or acquire them more slowly (Wilson et al. 2017; Ferrigno et al. 2020). Preschoolers and adults with no Western-style schooling perform somewhere in between monkeys and schooled adults (Deheane et al. 2022). The different capacities of schooled humans, non-schooled humans, and monkeys suggest that activities commonly performed by humans and reinforced in schools facilitate the efficient encoding of symmetries and nested repetitions. If I had to guess, I would guess that when subjects observe, produce, or manipulate regular sequences repeatedly, as Westerners do in school, their brains construct a compressed sensorimotor code for such sequences that relies on symmetries, repetitions, and nesting of operations, possibly in a way that is facilitated by acquiring a natural language. Note that there is independent evidence that nonhuman brains can memorize sequences in highly compressed ways (Liu et al. 2024).

The evidence presented by Dehaene and colleagues is consistent with (human) neurocognitive systems possessing one of at least five distinct algorithmic/architectural features, which I list here from least to most speculative: (i) the ability to encode compressed representations of symmetries and nested repetitions of operations nondigitally by using ordinary (compositional) neural representations (cf. Xie et al. 2022 for the non-compressed version of such representations); this hypothesis is consistent with the denial of LOT; (ii) the ability to encode and manipulate nonlinguistic-yet-compressed representations of sequences facilitated by acquiring a natural language; this hypothesis is consistent with a Nonclassical LOT; (iii) the ability to encode compressed representations of sequences digitally; such representations need not be especially language-like but this hypothesis goes at least part of the way towards Weak or Moderate Classical LOT; (iv) the ability to encode compressed representations of sequences in a language-like digital format; this hypothesis requires a Weak or, more likely, a Moderate Classical LOT; (v) the ability to encode compressed representations of sequences in the form of digital computer programs; this hypothesis entails Strong Classical LOT.

The issues raised by Dehaene et al. are complex and may point towards novel aspects of neural representation and computation. Yet their evidence does not entail that neurocognitive systems use Peircean symbols or any sort of language-like representations, let alone digitally encoded symbolic structures or computer programs. Therefore, their evidence does not unequivocally support LOT, let alone Classical LOT.

Our second example is Quilty-Dunn et al.'s (2023) detailed and sustained argument for LOT. I lack space for a detailed response, so the following brief observations will have to do.[32] The upshot will be that some of their evidence supports, at best, a generic (either Classical or Nonclassical) LOT. Perhaps this is all they intend, since it's unclear which version of LOT they endorse. In some passages, they imply they are defending the same view as Fodor, Pylyshyn, and other proponents of Classical LOT.[33] Furthermore, at least one of them defends Classical LOT explicitly elsewhere (e.g., Green and Quilty-Dunn 2021). Nevertheless, their official thesis is that *some* cognitive processes involve a LOT, and LOT representations form a natural kind with six properties: discrete constituents, role-filler independence, predicate-argument structure, logical operators, inferential promiscuity, and abstract conceptual content. This may sound like a version of Representational LOT (Chalmers 2023), to which they respond by endorsing Computational LOT explicitly (Quilty-Dunn et al. 2023, 72-3). Still, it remains uncertain whether they endorse Classical LOT or a generic version of Computational LOT. If they embrace the latter, or even better if they embrace Nonclassical LOT, I welcome them as allies. Still, it is instructive to see what their evidence does and does not support.

The main limitation of Quilty-Dunn et al.'s argument is that the six properties they use to define LOT, as they articulate them, fall short of a language-like representational format in the relevant sense. To be clear, I am *not* merely saying that each of their six properties is individually insufficient for a language-like format. I am saying that even *all* six properties, were they instantiated together, are insufficient for the relevant kind of format. Partly because of this, most of their evidence fails to support *any* LOT hypothesis properly so called, while some of it supports at best a generic (either Classical or Nonclassical) LOT. As I mentioned in Sect. 2, at a minimum, language-like representations should have constituents that play the semantic roles of linguistic subjects and predicates, which combine into structures that play the semantic roles of atomic sentences, which in turn combine into structures that play the semantic roles of complex sentences. Let's see how Quilty-Dunn et al.'s six properties compare.

Four of the properties listed by Quilty-Dunn et al.—discrete constituents, role-filler independence, inferential promiscuity, and abstract conceptual content—do not mandate a language-like format under any interpretation. Languages have these properties; many non-language-like representational systems have them too. Let's quickly see why. Depending on what is meant by "discrete", many non-language-like formats have discrete constituents; e.g., neural representations often cluster into discrete categories and composite iconic representations can be built out of discrete constituents (e.g., using pictograms). Role-filler independence can be exhibited by any system flexible enough to perform the same operation regardless of what its input represents (cf. Millière and Buckner 2025). Any representational format can be used in inference; degree of inferential promiscuity depends on factors that include aspects of computational architecture that go beyond representational format. Finally, many non-language-like formats can represent abstract contents; e.g., contemporary (Nonclassical) artificial neural networks (cf. Buckner 2023). Since these four properties do not discriminate between language-like and non-language-like formats, a fortiori they do not discriminate between Classical and Nonclassical LOT.[34] The remaining two properties require more careful treatment.

[32] A more detailed rebuttal to the sort of argument offered by Quilty-Dunn et al. 2023 with respect to perception may be found in Block 2023, though Block also appears to assume that cognition involves Classical LOT, without providing evidence.

[33] For example: "We will argue [that] … [i]n the half century since Fodor's (1975) foundational discussion, the case for the LoTH has only grown stronger over time" (1); "These properties [which define LOT] are intended to capture the spirit of earlier presentations of LoTH – a combinatorial, symbolic representational format that facilitates logical, structure-sensitive operations (Fodor & Pylyshyn, 1988)" (Quilty-Dunn et al. 2023, 3).

Predicate-argument structure is described by Quilty-Dunn et al. as "distinctively LoT-like" (2023, 3). This may or may not be true depending on what is meant by "predicate-argument structure". Without greater precision, attributing predicate-argument structure to a representational system is consistent with neurocognitive systems possessing one of at least four distinct algorithmic/architectural features, listed here from least to most speculative: (i) the ability to bind representations of properties to representations of objects, which does not require a language-like format and hence is compatible with rejecting any LOT hypothesis;[35] (ii) the ability to combine (possibly non-digital) representations semantically equivalent to linguistic predicates with representations equivalent to linguistic arguments (i.e., roughly, representations of sentential subjects), which is compatible with a Nonclassical LOT; (iii) the ability to combine digitally-encoded mentalese predicates with digitally-encoded mentalese arguments, which requires Weak or Moderate Classical LOT; and (iv) the ability to execute digitally-encoded instructions with a predicate-argument structure, which entails Strong Classical LOT. The main evidence the authors provide for predicate-argument structure is of two kinds: first, subjects can track objects while objects change some of their properties (Sect. 4.1.2), which is consistent with (i) and, therefore, the denial of LOT; second, the logical structure of linguistic input can affect implicit attitudes (Sect. 6.2), which is consistent with (ii) and thus with a generic (Nonclassical) LOT. In sum, some of their evidence for predicate-argument structure is neutral with respect to the LOT hypothesis, while the rest supports a generic (either Classical or Nonclassical) LOT.

[34] Hafri et al. "advance the case for LoT-like representation in perception" on the grounds that at least some perceptual representations exhibit "LoT properties: *Discrete constituents, role-filler independence, and abstract content*" (2023, 45, emphasis original). I agree that some perceptual representations are likely to have those properties. Pace Hafri et al., this reinforces the conclusion that such properties do not discriminate between language-like and non-language-like formats.

[35] Denying this trivializes the LOT hypothesis. Consider that Kazanina and Poeppel (2023) point out that paradigmatic neural representations such as hippocampal place cells, grid cells, boundary cells, head-direction cells, object cells, etc. can represent abstract properties and exhibit role-filler independence. This suggests that, once again, representing abstract properties and role-filler independence do not mandate a language-like format (cf., e.g., Frankland and Greene 2020; Schwartz and Fresco 2025). In addition, Kazanina and Poeppel maintain that grid cells and the like function as LOT *predicates*: "the neurobiological mechanisms found in the rodents' spatial navigation system are ontologically sufficient to represent symbols and operations required by the LoT" (p. 1007). If so, then ordinary neural representations are LOT predicates. If this were accepted, the result would be a radical and thoroughly Nonclassical version of LOT quite distant from LOT's original analogies with either natural or formal languages (cf. Sect. 4.3; van Bree 2024).

The last property, logical "operators", is also ambiguous between four possible algorithmic/architectural features, from least to most speculative: (i) the ability to perform logical or logic-like operations (e.g., equivalent to Boolean connectives or quantification), which even relatively simple, Nonclassical neural networks can have and is consistent with the denial of LOT; (ii) the ability to use (possibly non-digital) representations equivalent to logical connectives and quantifiers, which is compatible with a Nonclassical LOT; (iii) the ability to combine digitally-encoded logical connectives and quantifiers with digitally-encoded mentalese sentences to perform logical inferences on the sentences, which requires a Weak or, more likely, Moderate Classical LOT; and (iv) the ability to perform logical operations by executing digitally-encoded symbolic instructions, which entails Strong Classical LOT. The main evidence they provide is that humans can learn concepts that have Boolean or quantificational structure (Sect. 3 of their article; Piantadosi et al. 2016); and that humans and some nonhuman animals, when shown first a reward being hidden in one of two cups behind an occluder and then the empty cup, select the other cup (which contains the reward) without looking inside first (Sect 5.2 of their article); this suggests that humans and some nonhuman animals can perform inferences equivalent to Boolean negation and disjunction. Such evidence is consistent with (i), which is consistent with the denial of LOT. Thus, such evidence does not unambiguously support LOT, let alone Classical LOT.

Perhaps because they recognize some of these limitations, Quilty-Dunn et al. admit that "[m]any, perhaps all, of these properties are not necessary for a representational scheme to count as an LoT, and some may be shared with other formats" (ibid., 2). I agree and add that the main problem is lack of sufficiency: even all six properties (as they define them), collectively, are insufficient for a language-like representational format. Partly due to this, most of the evidence they provide does not clearly support *any* LOT hypothesis, and none supports Classical LOT.[36]

In conclusion, recent defenses of (Classical?) LOT hypotheses point at behavioral evidence that is mostly consistent with cognizers relying on representational formats that are *not* language-like, and even the evidence that suggests a language-like representational format is consistent with a Nonclassical LOT hypothesis. Thus, *no* empirical evidence supporting Classical LOT has been offered. Nevertheless, we've seen that there is compelling evidence for a generic Representational LOT in the form of processing outer speech and its neural signature, inner speech and its neural signature, and linguistic determinants of thought.

### 4.2 Evidence for Computational LOT

[36] Some of the peer commentaries go partway towards this conclusion in ways that complement my discussion (e.g., Attah and Machery 2023; Griffiths et al. 2023; Madva 2023; Pereplyotchik 2023, Roskies and Allen 2023).

As I mentioned, computation is the only known type of physical process capable of processing language-like structures in ways that match the syntactic, semantic, pragmatic, and inferential properties of linguistic structures. Because of this alone, if Representational LOT is true, then some form of computation is probably the way mentalese is processed. More broadly, there are a couple of reasons that at least some core neurocognitive processes are computational (cf. Piccinini 2020). Here I briefly present them in outline.

**The Argument from Medium Flexibility**

1. Rule-governed, medium-flexible functional mechanisms are computational
2. Neurocognitive architecture is (at least in part) a rule-governed, medium-flexible functional mechanism

----------------------------------------------------------------------------------------------------

Therefore, neurocognitive architecture is (at least in part) computational

A functional mechanism is a mechanism with teleofunctions (Garson 2013). A rule-governed mechanism is one that does not operate at random but in accordance with a rule, where a rule is a mapping from inputs and internal states to outputs. The first premise states that if a rule-governed mechanism has teleofunctions that can be realized in different media, then that is a computing mechanism.[37] It is a consequence of most accounts of physical computation, which see computation as rule-governed, teleofunctional, medium-flexible physical processes (cf. Anderson and Piccinini 2024).

The second premise expresses, in distilled form, the research program of much computational neuroscience (e.g., Dayan and Abbott 2005; Mallott 2024). Neurocognitive processes involve hundreds of types of neurons that form myriad structures and send electrical signals that release over a hundred types of neurotransmitters. Some of this complexity of structure and function is understood and much remains to be understood. Nevertheless, from this same complexity, computational neuroscientists infer structural and functional principles that do not depend on all biophysical details. These principles have to do with yielding the values of certain higher-level variables, such as output spikes or spike sequences, from the values of certain equally higher-level input variables under appropriate conditions, where the properties of the spikes that make a functional difference are properties such as frequency or timing, which are defined in ways that are largely independent of the physical media (voltages, ions, neurotransmitters) in which they are realized and, therefore, could be realized in other media. The relations between inputs, internal states, and outputs are not random but accord with a rule. An example of a rule is performing a rectified linear summation on inputs. These principles and relations can be captured mathematically and realized in computational models of neurocognitive functions.

[37] Medium flexibility should not be confused with multiple realizability, which pertains to functions that can be realized by manipulating the same medium in different ways. Medium flexibility is sometimes called “medium independence” or “substrate neutrality”, terms that have generated some confusion. Seth (2025) uses “medium flexibility” in the same sense, while Kirkpatrick (2022) uses it in a related sense.

The stunning success of artificial deep neural networks (DNNs) and the AI revolution they are bringing about (LeCun et al 2015) shows that there is something right about the neurocomputational approach to neurocognitive functions. Artificial DNNs have been able to match and sometimes surpass human cognitive capacities. Some DNNs, known as large language models, specialize in processing language-like structures, and their competence with human language vastly surpasses artificial systems with "symbolic" architectures consistent with Classical LOT (Khurana et al. 2023; Fang et al. 2024; Lappin 2024).[38] While there are many differences between (artificial) DNNs and neurocognitive systems, DNNs exhibit their capacities by reproducing at least some of the architectural features and computational principles discovered by neuroscientists (e.g., Cohen et al. 2022; Doerig et al. 2023) while realizing those same principles in artificial systems that are physically very different from neural tissues. Thus, DNNs provide further evidence for the second premise.

To the extent that large language models process language by possessing (sequences of) states that mirror linguistic structure, they also provide direct confirmation of Representational and Computational LOT. At any rate, processing language-like structures is a paradigmatic example of the kind of complex process that seems to require a computational architecture. Thus, insofar as brains process language-like structures (Representational LOT), it is likely that they do so by possessing an appropriate computational architecture (Computational LOT). In conclusion, the success of computational neuroscience and DNNs support Computational LOT, although a more adequate argument would require a more detailed canvassing of the literature than I have space for.

**The Argument from Complex Information Processing**

1. Sufficiently complex information processing requires computation
2. Neurocognitive systems process information in sufficiently complex ways

-----------------------------------------------------------------------------------------------------

Therefore, relevant neurocognitive processes are computational

This argument begins with the observation that computation is the only known physical process capable of processing information-bearing states in ways that match their semantic properties. Cognition seems to involve the processing of information-bearing states in ways that match their semantic properties, and neurocognitive systems process information in ways that are as complex as any. Processing language-like structures in ways that match their semantic properties is a paradigmatic example of complex information processing, so this argument applies to Representational LOT. Variants of this argument can be found in all corners of the mind sciences. This is one reason that the idea that cognition involves computation dominates the mind sciences (Colombo and Piccinini 2023).

[38] Here are four ways in which large language models are Nonclassical: (1) they encode information nondigitally as values along multidimensional continuous scales, (2) they encode language not by encoding their syntactic structure explicitly but by compressing statistical structure across multiple linguistic scales that go beyond syntax to include sub-lexical patterns, semantic relations, and discourse and topic structure; (3) since they encode information in ways that are multidimensional and continuous, by necessity they process such information using nondigital operations (using matrix multiplication; attention weights; nonlinear activations; residual connections); (4) they learn by associative learning rather than explicit hypothesis testing.

Both arguments make a compelling case that, if neurocognitive systems manipulate language-like structures, and particularly if they do so in ways that match their semantic properties (i.e., Representational LOT), then they do so by computing over language-like structures (i.e., Computational LOT). Recall, however, that computation need not be digital and the neurocognitive architecture need not be Classical, so Classical LOT does not follow from any of this. If Classical LOT is to hold, it requires independent support.

### 4.3 Evidence for and against Classical LOT

I will now argue that empirical evidence accumulated over the past several decades is so overwhelmingly against Classical LOT that Classical LOT is no longer a viable hypothesis.

For starters, recall that Weak Classical LOT simply states that language-like neural representations are digitally encoded and processed by digital computing systems. This does not give us a lot of explanatory power without at least the further assumption that there are digital memories, distinct from processors, where mentalese data structures can be stored, and this yields Moderate Classical LOT. If we add the further assumption that the processors execute programs stored in memory, we reach Strong Classical LOT.

In Sect. 2, I mentioned three kinds of digital computing systems: hardwired, plastic, and program controlled. Program control provides a lot of computational flexibility at the cost of a very specific and delicate control structure. Aside from program control, digital circuits can be hardwired or plastic. As I already mentioned, neural circuits are usually plastic, so let’s set aside hardwired circuits. Given that I am allowing digital circuits to be plastic, the Classical-Nonclassical dichotomy turns into a spectrum with a grey area in between. It’s worth briefly discussing which portions of the spectrum we are primarily interested in.

On the Classical end of the spectrum are hardwired digital networks of logic gates like those that make up conventional digital computers. On the Nonclassical end of the spectrum are nondigital, plastic networks of neurons—that is, networks that encode information nondigitally and thus, a fortiori, compute by means of nondigital operations. If we travel along the spectrum from the Classical end towards the Nonclassical end, in the middle we find networks that encode information digitally and can be trained to develop a digital-step-by-digital-step architecture (cf. Turing 1948; Copeland and Proudfood 1996) and then networks that encode inputs and outputs digitally but can perform nondigital operations (e.g., because they process signals continuously, or encode intermediate steps nondigitally, or because their computations are holistic or subsymbolic; cf. Aydede 1997; Smolensky and Legendre 2006; Kleyko et al. 2022, 2023).

Much of the Classicism-Connectionism debate from the 1980s and 1990s occurred within this grey area of networks that encode inputs and outputs (approximately) digitally but are plastic and perform nondigital intermediate operations. Sometimes Classicists argued that such networks are either insufficiently Classical or, if they are sufficiently Classical, they are implementations of Classical systems (e.g., Fodor and Pylyshyn 1988) while Connectionists argued that such networks are sufficiently Nonclassical to count as alternatives to Classical LOT (e.g., Smolensky 1988). More recently, Papadimitriou et al. (2020) developed an idealized, neurally inspired model—the assembly calculus—showing in principle how networks of spiking neurons could manipulate neuronal assemblies to implement Turing-complete computations.

We don't need to resolve the Classicism-Connectionism dispute because these "Connectionist" models in the middle of the spectrum are not realistic neurocomputational models. They are motivated primarily by engineering, behavioral, or mathematical considerations. Even where they are loosely inspired by neural principles, many of their mechanisms and operations (e.g., binding via circular convolution, permutation, or large-scale connectivity patterns) are unsupported by current neurophysiology and do not correspond to any known neural processes or circuits.[39] I am not defending this kind of "Connectionism". I am defending the view that the computational architecture of the brain can only be discovered by studying brains empirically and by building models that, unlike both Classical and many Connectionist models, are constrained by evidence about how brains work. As I will argue presently, brains encode and process information nondigitally and lack most of the features of Classical architectures. Therefore, whether we choose to classify neural networks that fall in the grey area between the Classical and Nonclassical ends of the spectrum as Classical or Nonclassical does not matter much.

Three types of reasons have been given to support Classical LOT: behavioral evidence of certain cognitive capacities (e.g., language processing, or the evidence reviewed by Dehaene et al. 2022 and Quilty-Dunn et al. 2023), architectural arguments to the effect that a computational architecture that supports at least Moderate Classical LOT is the only possible explanation of such cognitive capacities (e.g., Fodor and Pylyshyn 1988; Gallistel and King 2009), and Classical computational models of cognitive capacities.

In Sects. 4.1 and 4.2, I argued that the available behavioral evidence strongly supports Representational LOT and Computational LOT, but it does *not* support Classical LOT. In Sect. 3, I argued that what the architectural arguments support is merely the conditional that, *if* the Classical LOT hypothesis holds, *then* brains must contain the relevant computational architecture. So, the question remains, do brains contain the kind of computational architecture that is required for Classical LOT to hold?

[39] A possible exception: as I pointed out in Sect. 3, humans can follow any sequence of linguistic instructions they can memorize on any data they can memorize, thereby becoming equivalent to computationally universal systems until they run out of memory. Under such special constraints, perhaps some neural systems operate in ways that approximate digital computation at a coarse level of organization.

Before we look at relevant empirical evidence, I will say a few words about Classical computational models of cognition. Classical models, which purport to explain cognitive capacities in terms of symbolic (digital) computations, form a storied tradition that includes General Problem Solver (Newell and Simon 1972), ACT-R (Anderson 1983), and SOAR (Newell 1990), among others. The point I made about architectural arguments—that they fail to support the Classical LOT hypothesis due to a combination of underdetermination and computability considerations—applies to Classical computational models as well. Some coarse-grained features of neurocomputational systems can be successfully inferred from some combination of behavioral capacities and models that capture such capacities. But the format of neural representations and the details of neurocomputational architecture cannot be inferred from behavior and behavior-based models alone.

The same function can be computed by indefinitely many algorithms, some fully digital and Classical (of which some are hardwired, others are based on program execution), some in the grey area between Classical and Nonclassical, many nondigital (and hence Nonclassical), some known, and many unknown. Therefore, in the absence of neuroscientific evidence that brains have the computational architecture that supports the algorithms and representations posited by a model, the mere success of a computational model, no matter how well it matches behavioral evidence (including error rates and reaction times), cannot by itself support any precise hypothesis about computational architecture, representations, and algorithms.

Given the evidence that brains have a Nonclassical architecture (to be reviewed presently), many former proponents of Classical models have embraced the cognitive neuroscience revolution and transitioned towards Nonclassical models or at least the view that Classical or quasi-Classical models are just rough approximations of a Nonclassical brain (Boone and Piccinini 2016, 1529-30). Yet Classical models and their successes are still sometimes mentioned as putative evidence that the brain itself is Classical. A case in point is the so-called Probabilistic LOT (PLOT) family of models recently developed by Tenenbaum and associates (Griffiths et al. 2024). The main innovation of PLOT models is that, unlike typical Classical LOT models, PLOT algorithms compute over representations of probabilities in accordance with Bayes' theorem.

As PLOT proponents point out, the kind of Bayesian inference they posit is computationally intractable—that is, it requires more representational and computational resources than brains can muster. Therefore, brains cannot literally work in the way described by PLOT models. Instead, PLOT proponents suggest that brains *approximate* PLOT models (e.g., Vul et al. 2014). This suggestion may be interpreted in two very different ways.

One interpretation is that brains have a Classical architecture that employs some computationally tractable heuristic(s) that sometimes approximate(s) Bayesian inference. This is consistent with the Classical LOT hypothesis.[40] To make it plausible, there needs to be empirical evidence that brains have a Classical architecture plus evidence that the hypothesized Classical neural computations follow heuristics that approximate Bayesian inference. If such evidence were found, Classical LOT would be vindicated. On the contrary, we will see shortly that there is overwhelming evidence that brains have a Nonclassical architecture.

[40] This interpretation is suggested by statements to the effect that cognition literally involves mentalese representations employed by PLOT models. For instance:

The other interpretation of PLOT models is that brains have a Nonclassical architecture that, in relevant cases, computes in ways that approximate Bayesian inference without relying on digitally encoded symbols or symbolic structures (Rescorla 2023; cf. Griffiths et al. 2024, Ch. 12). This is plausible, can be assessed on a case-by-case basis, and does nothing to vindicate Classical LOT, or even LOT simpliciter. Here it's worth adding that, if neurocognitive systems approximate Bayesian inference Nonclassically, cognition is better explained in terms of the architecture, representations, and algorithms by which they do so (cf. van Roij et al. 2012; Craver and Kaplan 2020; Piccinini 2020a; Griffiths et al. 2023).

We are finally ready to briefly sample some important findings about neurocognitive systems that bear on their computational architecture and militate against Classical LOT (cf. any textbook on computational neuroscience, such as Dayan and Abbott 2005; or Mallott 2024):

*Representational format*. Some neurons transmit graded potentials, which bear no resemblance to digital signals. In contrast, the most typical signals exchanged between neurons are spikes or action potentials, which are released all at once with a certain probability when excitation within a neuron reaches a threshold. The probability of action potentials depends on several factors including the stochasticity of ion channels and various modulatory and metabolic effects. Still, when they occur, spikes are all-or-none, which led McCulloch and Pitts (1943) to model them as if they were digital. McCulloch and Pitts's model of the brain as a digital computing system was probably the biggest historical influence on the origin of the Classical LOT hypothesis. Be that as it may, it's been known for a long time that, as McCulloch and Pitts themselves realized, action potentials are not digital in the sense needed to encode information digitally. Piccinini and Bahar have defended this point at length elsewhere (Piccinini and Bahar 2013, revised as Ch. 13 of Piccinini 2020a) and I am not going repeat their full analysis here. In brief, here are a few reasons: unlike digital signals, spikes do not occur within well-defined finite time intervals; spikes from distinct neurons may be more or less synchronous in a graded way that appears to contribute to neurocognitive functioning but are not synchronized in the precise way needed to form a digital code; and the functionally relevant properties of spikes are

---

> According to the theory, our knowledge of the world is organized into concepts that we combine in language-like ways. The content of a concept is a function or subroutine in a probabilistic programming language; when faced with a new situation, we draw on a rich library of these concept building blocks to compose an appropriate model of the situation on the fly, much as a programmer might code up a script in Python. The resulting model—a program in the probabilistic language of thought—encodes a probability distribution over world-states that is sufficiently precise to reason in combinatorial ways. (Lew et al. 2020, 1)

These authors' analogy between concepts and computer programs could be interpreted as a version of Strong Classical LOT.

frequency and timing, both of which are graded rather than digital. In part due to lack of precise synchronization and in part because spike frequency and timing are the relevant variables, spikes are not concatenated into well-defined digital strings; they are too stochastic in ways that cannot be categorized as probabilistic digital states; and spiking is subject to many graded (and hence non-digital) modulatory effects. As I pointed out, higher-level neural representations can cluster into discrete categories—but they remain graded and composed of (graded) spike trains, and they are not processed via primitive digital operations defined over atomic symbols (as Classical LOT requires) but via nondigital neural computation that operates on entire hierarchies of features that do not resemble semantically atomic symbols (more on this below). This lack of a digital representational format undermines all versions of Classical LOT.

*Encoding scheme*. Digital encoding relies on two properties of digital vehicles: finitely many digit types and positioning within a string. For instance, decimal numerical codes require ten types of digits (0, 1, … 9) and unambiguous positioning of the digits within a string: the rightmost digit represents units, the second digit to the left represents tens, and so forth. Neural systems do encode information, in part, based on where vehicles occur within the system, which might suggest a similarity with the positioning of digits within strings. For instance, adjacent cortical columns within mammalian visual area V1 encode information from adjacent (visual) receptive fields. But this so-called *place code* is not a digital encoding scheme. For one thing, activity within distinct cortical columns is graded and thus does not fall into finitely many types. Equally importantly, cortical columns and other components of nervous systems blend into one another and so are not sharply demarcated in the way that digital components that carry distinct digits need to be. Most significantly, neural systems encode information not only by relying on physical location but also, and primarily, by using features that include the strength of connections between neurons, firing rates and timing, patterns of activation within neuronal populations, and dynamical evolution of the activation patterns. I've already pointed out that firing rates and timing are graded and hence not digital. Thus, they are not digital codes.[41] Similar points apply to neural connection strengths, patterns of activation within neuronal populations, and dynamical evolutions of activation patterns. This lack of digital encoding schemes undermines all versions of Classical LOT.

*Neural computation*. Since neurons do not encode information digitally, a fortiori they do not perform digital operations. Instead, typical neurons combine two main types of computational operations: first, their dendrites and soma integrate the many inputs they receive (which are typically fairly discrete when received individually but are then integrated with inputs from thousands of other synapses received over continuous time), and then, if a certain activation threshold is reached, the soma and axon generate and transmit action potentials to other cells. As I have pointed out, neither of these operations are strictly digital. Further disanalogies between neurons and digital computing components include that neurons can not only excite but also inhibit one another and that neurocomputational operations are subject to many modulating factors. In complex neurocomputational systems, which process complex stimuli and guide complex behavior, the combination of (non-digital) information encoding and (non-digital) operations results in complex recurrent hierarchies of layers of neuronal populations representing and processing features of the relevant targets at many levels of abstraction, with extensive feedback between the different layers (cf. Ritchie and Piccinini 2024). Another disanalogy with digital systems is that, as far as we can tell, all the represented features of a target, whether "symbolic" (i.e., analogous to words) or "subsymbolic" (i.e., any other features of the target), can be relevant to processing neural representations. While no one knows precisely how neurocomputational systems process language-like representations, there is no reason to suspect that they do so by anything other than an extension of the same representational and computational strategies that they employ everywhere else—that is, by hierarchies of recurrent neuronal populations encoding and processing hierarchies of features of the relevant targets (in this case, language-like stimuli and responses, whether real or imagined) at different levels of abstraction without being limited to primitive syntactic operations on words in the way that Classical LOT systems are hypothesized to be. This is just a bare sketch of a hypothesis about how neurocognitive systems might process language-like representations, and there are plenty of controversial or unknown aspects of neural computation. But there is no evidence and no need to suppose that when processing language-like representations, brains suddenly turn to something as different from their usual modus operandi as digital representations and computations, especially since our best artificial systems for processing language-like representations work in ways that are very different from Classical systems.[42]

[41] During the 1990s there was a research program searching for repeating triplets or quadruplets of spikes that might encode information. In principle, if such precise spike patterns had been there, nervous systems might have used them to construct digital codes. For better or worse, they turned out to be statistical artifacts (Oram et al. 1999).

At this point, a proponent of Classical LOT might wonder whether neural signals and operations that are nondigital at the level of single neurons or small neuronal populations might constitute digital codes at a higher level of organization. After all, all kinds of physical states can be coaxed into digital codes, including neuronal assemblies (Papadimitriou et al. 2020). The problem is that coaxing physical states into digital codes, and state transitions into digital operations, requires an enormous amount of careful regimentation of the right sort. Ordinary digital computers accomplish this by heroic engineering feats that include very precise placement of the components, very precise digital clocks to synchronize components, and very sophisticated control systems to coordinate the actions of the components. The result is that only select states at select times count as (digital) computational states, while the system ignores all the irrelevant states. Because of all this regimentation, those select states and transitions between them bear a physical signature of the digital computations they implement (Anderson and Piccinini 2024). Neural representations and operations do compose higher-level representations and operations, but the evidence we have suggests that they retain the same sort of graded (and hence non-digital), logarithmically scaled properties exhibited by spike trains (cf. Buzsáki 2019, Ch. 12). Worse, there is no evidence of the sort of precise and exact placement, synchronization, and control mechanisms that are needed to implement digital encoding schemes and operations. On the contrary, nervous systems exhibit an organizational structure known as *small-world network* (Watts and Strogatz 1998) that is often *scale-free* (Barabási and Albert 1999). A small-world network is a network in which the nodes are neither randomly connected nor connected in a highly regular way; they are connected through a mixture of randomness and order that results in a high degree of clustering and a relatively short path between any two nodes. Recent studies show that small-world network properties are crucial for efficient neural computation and are linked to various neurological conditions (e.g., Hagan et al. 2025; Palma-Espinosa 2025; Wu et al. 2025). In contrast, digital computation requires very precise (nonrandom) connectivity patterns between the nodes. Neurobiological systems exhibit many other experimentally established properties that are radically different from those needed to support digital (and hence Classical) computation, but I hope I've provided enough examples for present purposes. As a final point, note that network science, the formal discipline that has been yielding insights into the organization and control structure of neurocomputational systems (e.g., Faskowitz et al. 2022), came into existence decades later than the Classical LOT hypothesis.

[42] Compare footnote 38.

*Modularity*. The brain is not as modular as needed to support an explanatorily adequate Classical architecture. The Weak Classical LOT hypothesis is consistent with brains being one giant, undifferentiated FSA. But this is explanatorily weak. When it comes to Moderate and Strong Classical LOT, the posited architectures separate processors from memory components and, typically, divide the cognitive labor among many distinct modules. Each module has its own processor(s) and memory for storing data and, perhaps, programs (if it's a program-controlled module). The separation between modules as well as between processors and memory is needed for each processor to perform well-defined digital operations on its inputs (of well-defined size), possibly in response to (well-defined, digital) control signals. Thus, any explanatorily adequate Classical LOT architecture requires that computing and memory functions be localized within appropriately segregated neural structures. For better or worse, there is increasing evidence that neurocomputational systems are not quite modular in that way—or equivalently, that cognitive functions are not localized in neural structures in the requisite way. The degree of departure from strict modularity is disputed and there are a range of options under discussion, and I lack space to do justice to this topic. Suffice it to say that neural structures can often participate in many cognitive functions (neural reuse), many structures can perform the same functions (degeneracy), and, most relevantly, the computational operations performed by neural structures are subject to graded (and hence nondigital) modulatory effects that depend on the task the organism is engaged in (cf. McCaffrey 2023; Pessoa 2022; Westlin et al. 2023; Zerilli 2021).

*Memory*. Memory within neurocomputational systems does not store digitally encoded symbols and does not function in the way required by a Classical architecture. This point requires some elaboration. Neural systems have several means of preserving information—i.e., memories—at different spatiotemporal scales. The best-known are (1) keeping neurons or neuronal populations active for some time, possibly in states that are among the dynamical attractors of a neural network that remains in that state until a new input nudges it out of it; (2) generating and spreading waves of activity through a neural system; (3) altering the strength of existing active connections (synapses) between neurons within a specific population so that relevant patterns of neural activity are generated under appropriate conditions; (4) deactivating active synapses or activating dormant ones; (5) making new connections (synaptogenesis), and (6) generating new neurons (neurogenesis).

These types of memory occur within neuronal populations that play three intertwined roles at once: performing ordinary cognitive functions (processing sensory information while guiding behavior), learning, and memory. Typical neuronal populations cannot store arbitrary information—they can only store information for which they specialize in ways that depend on their connections with specific sensory systems, motor systems, or other neural systems. There is no known type of neuronal memory that functions like a digital memory, which is a system of distinct cell arrays with distinct and precise locations (or even better, addresses with digitally encoded names) that can be called upon as needed to retrieve specific and arbitrary digital states and that are wholly distinct from processing components.

The dominant view of long-term memory (LTM) is that LTM is stored primarily in patterns of synaptic structures that connect neurons. While synaptic changes along with their support and modulation by glial cells are a huge part of the story (Ortega-de San Luis and Ryan 2022), I doubt that synaptic changes are the whole story. I doubt it because there is evidence that synaptic structures are not stable for as long as organisms retain their memories, LTM survives the disruption of synaptic connections (e.g., Chen et al. 2014; Ryan et al. 2015), neurons can communicate by transferring proteins and RNA via exosomes (Smalheiser 2007), simple forms of memory (habituation, sensitization, classical conditioning) can be transferred from one organism to another by transferring RNA molecules (Bédécarrats et al. 2018), some simple forms of memory can be transmitted to offspring epigenetically (Dias and Ressler 2014), Purkinje cells can alter the timing of their response (Johansson et al. 2014; Jirenhed et al. 2017), and even single, non-neural cells exhibit aspects of memory (Kukushkin et al. 2024). This evidence suggests that neurons might encode some information by means of molecular structures internal to them, such as RNA molecules or epigenetic changes (e.g., Kyrke-Smith and Williams 2018; Griffith et al 2024).

RNA and DNA have digital structure, which could be used to encode information in a digital format. If that is the case, it's important to understand what sort of information neurons might encode in this way. The evidence I just listed suggests that molecules within neurons might encode information about which genes should be expressed under certain conditions or, if neurons have molecular signatures that they can communicate, which other neurons a neuron should connect and communicate with synaptically, and perhaps the strength of such connections or the type of communication. Even if this kind of information is stored molecularly inside neurons, LTM will continue to involve making, altering, and restoring connections between neurons within a population so that relevant patterns of neural activity are generated under appropriate conditions; what might be different is the means—molecules inside the cell rather than just synaptic structures—by which neurons retain information about such connections in the long run. In other words, the roles plausibly played by internal molecules in memory are compatible with mainstream (Nonclassical) neurocomputational explanations (cf. Gold and Glanzman 2021; Gershman 2023; Colaço and Najenson 2023). Thus, even if some aspects of LTM involve molecular encoding of information within neurons, this does not suggest that the kind of memory needed by the Classical LOT hypothesis exists anywhere in nervous systems.

This is an important point because in recent years, some have speculated that a kind of symbolic memory might exist within neurons in the form of RNA, and this might be a component of a digital nano-computer inside each neuron (Gallistel and King 2009; Gallistel and Balsam 2014; Gallistel 2021). Akhlaghpour (2022) has even shown that RNA *could* be used for universal digital computation by combining operations similar to those that occur within cells. That's very cool. Just because RNA could be used in this way, however, it doesn't follow that the brain so uses it. Anything with enough degrees of freedom, suitably regimented and organized, *could* be used for digital computation; it doesn't follow that any particular physical system is a digital computing system (cf. Anderson and Piccinini 2024). Or compare McCulloch and Pitts's (1943) demonstration that extremely idealized and simplified neurons *could* be used to construct Boolean circuits and finite state automata (Kleene 1956), which was adopted by von Neumann (1945) to describe the design of artificial general-purpose digital computers. Just because something somewhat like neurons *could* be used in this way, it doesn't follow that the brain uses actual biological neurons in this way. In fact, there is no evidence that brains use neurons to build digital circuits and, as I argued above, the evidence we have suggests that they don't. RNA is surely involved in cognition, if nothing else because gene expression is involved in memory consolidation and gene expression requires RNA. In addition, as I said above, RNA or some other molecules inside neurons might encode information about which other neurons a given neuron should connect and communicate with. That is a far cry from the notion that neurons contain digital nano-computers made of RNA.

Three more points are worth making about hypothetical RNA-based digital nano-computers. First, to build a functioning digital computer, it's not enough to have a digital code such as RNA molecules; it takes a lot of careful regimentation and control and there is no evidence of that. Second, spike trains are the primary signals through which neurons communicate with each other and drive muscle contractions. If there were digital nano-computers inside cells, their inputs and outputs would have to be transduced from and into spike trains, respectively. Third and finally, virtually all the empirical evidence we have suggests that the vehicles or our thoughts, and the drivers of behavior, are signals (mostly, spike trains) from neuronal populations supported and modulated by glial cells. As I've said, spike trains are not digital and are not computed digitally. Therefore, *even if* there were digital nano-computers inside neurons, this would not really support Classical LOT, because to be used by the brain to drive thought and behavior, the outputs of such hypothetical digital nano-computers would have to be transduced into spike trains, which are not digital and, with the Nonclassical exceptions I discussed, are not language-like. At any rate, there is no evidence of digital nano-computers inside neurons and no reason to posit them.

*Situated learning*. Digital computing systems increase their computing power either by increasing the complexity of their control systems or by adding memory storage, within memory components that are structurally separate from the processing components. If any learning occurs, it consists of either altering programs stored in memory or, perhaps, altering circuits so that they will compute a different function in the future. Either way, learning and computing functions are performed at different times. In contrast, typical neurocomputational systems do triple duty as processing devices, memory devices, *and* learning devices at the same time. As they process information, neurocomputational systems rely on information stored in the connections between their units while also learning how to improve their future information processing by altering those same connections between their units in a graded way. This is incompatible with the sharp separation between processors and memory, and between information processing and learning, that digital systems require to perform their operations correctly. This incompatibility is a robust critique of the analogy between neural and digital systems that goes back to von Neumann (1958). In addition, the integration of processing, learning, and memory requires a degree of situatedness within the body and environment that is missing at least from conventional digital computers. This situated integration of processing, learning, and memory allows neurocognitive systems to learn to build their own representations *while* learning to process them. This same situatedness of neural computation and representation solves a chronic problem that Classical LOT theorists had been unable to solve: how, in the absence of a programmer, do biological computations acquire the ability to process representations in accordance with their content? Piccinini (2022, 2024) has argued that situated Nonclassical architectures provide a solution: neurocognitive systems are inherently situated in a sense in which (conventional) digital systems are not, and they build their own representations via development and learning by both integrating processing, learning, and memory *and* by receiving extensive and continuous feedback from their body and environment in ways that require accounting for their own movements through efferent copies of their motor commands (cf. Buzsáki 2019). In fact, the semantic content of neural representations is itself a function of the (inherently situated and Nonclassical) way in which they are constructed. Brains' lack of digital memory, of sharp separation between memory and processing, and of sharp temporal separation between learning and processing, along with their situated integration of processing, learning, and memory undermine both Moderate and Strong Classical LOT.

To sum up, Classical LOT requires a digital code for the language-like data (and programs, in the case of Strong Classical LOT) plus the relevant hardware: digital processors, digital memory separate from the processors to store data (Moderate and Strong Classical LOT) and instructions (Strong Classical LOT), and possibly specialized control systems for decoding and executing programs (Strong Classical LOT). Any digital computing system with memory separate from the processors needs specialized digital devices to keep track of the memory locations, fetch the right data stored in memory, keep track of intermediate results, and store new data in memory. If it stores and executes programs, it also needs control structures that decode instructions and select the right operations, program counters, and other control structures to track which instruction needs to be executed next. Needless to say, the computing machinery Classical LOT requires goes beyond the simplistic digital model of neural networks that McCulloch and Pitts (1943) proposed, which even at that time was a gross simplification and idealization of biological neural networks and which no one considers relevant to understanding neurocognitive systems.

What we have learned since 1943 reinforces the conclusion that virtually none of the core architectural features required by Classical LOT are present within nervous systems. There is evidence that the specialized neural representations involved in explaining human linguistic cognition mirror some aspects of language-like structure at a coarse level of granularity and suggest a Nonclassical LOT of relatively modest scope, but there is no evidence of a genuinely digital code in the brain, or a computer-like programming language being executed within the brain, let alone digital processors, digital control structures, and digital memory separate from the processors. A fortiori, brains have a Nonclassical computational architecture and, therefore, Classical LOT ceased long ago to be an empirically viable hypothesis.[43]

## 5. Conclusion

I have argued that neuroscientific evidence about the computational architecture of the brain rules out Classical LOT in favor of a Nonclassical LOT hypothesis. Classical LOT theorists have attempted to defend their hypothesis based on behavioral evidence, without taking seriously evidence about neurocomputational architecture. This was a mistake because behavioral evidence is compatible with both Classical and Nonclassical computational architectures.

I have *not* defended empiricism, associationism, or Connectionism—views that are often contrasted with the Classical LOT hypothesis. Those views have merits but the degree to which they are correct is irrelevant to our present concerns. What I have argued is that the version of LOT that is empirically supported boils down to the following relatively modest theses:

(A) Biological brains represent and compute nondigitally, and hence Nonclassically.
(B) At least the human version of such a Nonclassical architecture has the capacity for cognitive processes that support and are supported by the processing of natural language.
(C) Some neural representations involved in the capacities mentioned in (B) mirror some of the structure of natural language and represent in a language-like way.

---

[43] An anonymous reader objected that we may not know enough about neurocomputational architecture to rule out the existence of a Classical architecture hidden at some level of organization. They refer to Jonas and Kording (2017), who argue that some current neuroscience methods may be too weak to reverse engineer a classical cognitive architecture from neural data. Their study applies (some) standard neuroscientific analyses—such as lesion studies, spike analysis, and functional connectivity—to a classical microprocessor (the MOS 6502). The processor's functional organization is fully known. Yet these methods failed to recover its hierarchical and modular architecture. The experiment builds on earlier work by James et al. (2010), who *did* reverse engineer the 6502's transistor-level activity by using more fine-grained methods analogous to contemporary connectomics techniques. Thus, James et al.'s work shows that with patience and powerful enough methods, we *can* reverse engineer a Classical architecture. At any rate, even if some of the tools of modern neuroscience were inadequate to reveal a Classical architecture at some scales, assuming such an architecture requires evidence in the first place. As I have argued, behavior can be explained without invoking Classical architectures, and there is plenty of neuroscientific evidence supporting a Nonclassical architecture. Until evidence of a Classical architecture is produced, we have *no* reason to believe a Classical architecture lies hidden in the brain.

Importantly, this Nonclassical LOT does not require a mentalese distinct from natural language, let alone one that is analogous to the formal languages of logic and computer science. All it requires is neural representations that have language-like structure at least at a coarse level of grain. Such neural representations are categorical, compositional, and can represent amodally. They encode information nondigitally and are processed by ordinary (nondigital, and hence Nonclassical) neural computations that rely not only on syntactic structure but many other features. They, along with neural representations involved in processing other public symbolic systems, are the only neural representations that are symbolic in Peirce's sense. To understand the neural representations that subserve language and discursive thought more deeply, we need to understand them at a finer level of grain rather than merely by analogy with natural language (cf. Coelho Mollo and Vernazzani 2024), and we need to understand how they emerge from underlying neural representations and computations.

There are many other considerations that militate against Classical LOT: graceful vs brittle degradation of performance under damage to neural circuits, evolvability through natural selection, embodiment, embeddedness, enaction, the entanglement of cognition and affect, and perhaps the role of consciousness in cognition. I purposefully mostly ignored those considerations in favor of an argument based on computational architecture. I did this because computational architecture is where Classical LOT has historically been claimed to have an advantage. I have argued that such a putative advantage was an illusion all along and, in any case, brains' computational architecture is Nonclassical. Anyone who appreciates other reasons to reject Classical LOT should be glad to see Classical LOT refuted on its own merits and should be reassured that the Nonclassical LOT I have defended is not only compatible with the situatedness of cognition but actually requires it (Piccinini 2022, 2024). It is also compatible with the evolvability of neurocognitive systems through natural selection, the entanglement of cognition and affect, and a role of consciousness in cognition.

As to Classical "symbolic" models of cognition, successful ones may still play a role as rough approximations of some cognitive processes at a coarse level of grain, without any implications about the representational formats, algorithms, or computational architecture that carry out those processes. Needless to say, this is not what the Classical LOT hypothesis says or how typical Classicists have intended their theory to be interpreted. And given the confusion Classical models have generated over the years, and the persistent tendency of many to overinterpret them as putative evidence that cognition itself is Classical, it would be safer to replace Classical models with Nonclassical ones.

In conclusion, Classical LOT is empirically ruled out by ample neuroscientific evidence. In contrast, Nonclassical LOT is a plausible part of the story of how humans think and acquire, process, and have their thinking enhanced by natural language.